%% file: HM_MG.tex
\documentclass[%
 reprint,
nofootinbib,
 amsmath,amssymb,
 aps,
]{revtex4-1}

\usepackage{graphicx}
\usepackage{dcolumn}
\usepackage{bm}
\usepackage{hyperref}

\usepackage{aas_macros}
\usepackage[capitalize]{cleveref}
\usepackage[dvipsnames,table]{xcolor}
\hypersetup{urlcolor=RedViolet,
	    citecolor=ForestGreen ,
	    linkcolor=RoyalBlue, 
	    colorlinks=true}
\usepackage[normalem]{ulem}

\def\simlt{\stackrel{<}{{}_\sim}}
\def\simgt{\stackrel{>}{{}_\sim}}
\newcommand{\Mpch}{$\mbox{Mpc}/h$}
\newcommand{\kpch}{$\mbox{kpc}/h$}
\newcommand{\msunh}{M$_{\odot}/h$}
\newcommand{\MpchInv}{$h/\,\mbox{Mpc}$}

\newcommand{\ie}{\textit{i.e.}}
\newcommand{\eg}{\textit{e.g.}}

\newcommand{\elephant}{\textsc{elephant}}
\newcommand{\halofit}{\textsc{halofit}}
\newcommand{\hmcode}{\textsc{hmcode}}
\newcommand{\lcdm}{$\mathrm{\Lambda \text{CDM}}$}
\newcommand{\fsig}{F(\sigma)}
\newcommand{\pk}{$P(k)$}
\newcommand{\pklin}{$P(k)_{\text{lin}}$}
\newcommand{\pkhm}{$P(k)_{\text{HM}}$}
\newcommand{\pksim}{$P(k)_{\text{sim}}$}

\newcommand{\pkmg}{$P(k)_{\text{MG}}$}

\begin{document}


\title{Improved analytical modeling of the non-linear power spectrum in modified gravity cosmologies}

\author{Suhani Gupta}
\email{[gupta, hellwing, bilicki]@cft.edu.pl}
\author{Wojciech A. Hellwing}%
\author{Maciej Bilicki}%
 
\affiliation{Center for Theoretical Physics, Polish Academy of Sciences, Al. Lotnik\'ow 32/46, 02-668 Warsaw, Poland}%

\date{\today}

\begin{abstract}

Reliable analytical modeling of the non-linear power spectrum (PS) of matter perturbations is among the chief pre-requisites for cosmological analyses from the largest sky surveys. This is especially true for the models that extend the standard general-relativity paradigm by adding the fifth force, where numerical simulations can be prohibitively expensive. Here we present a method for building accurate PS models for two modified gravity (MG) variants: namely the Hu-Sawicki $f(R)$, and the normal branch of the Dvali-Gabadadze-Porrati (nDGP) braneworld.  We start by modifying the standard halo model (HM) with respect to the baseline Lambda-Cold-Dark-Matter (\lcdm) scenario, by using the HM components with specific MG extensions. We find that our \pkhm{} retains 5\% accuracy only up to mildly non-linear scales ($k \lesssim 0.3$ \MpchInv) when compared to PS from numerical simulations. At the same time, our HM prescription much more accurately captures the ratio $\Upsilon(k) = P(k)_{\text{MG}}/P(k)_{\Lambda \text{CDM}}$ up to non-linear scales. We show that using HM-derived $\Upsilon(k)$ together with a viable non-linear \lcdm{} \pk{} prescription (such as \halofit{}), we render a much better and more accurate  PS predictions in MG. The new approach yields considerably improved performance, with modeled $P(k)_{\text{MG}}$ being now accurate to within 5\% all the way to non-linear scales of $k \lesssim 2.5-3$ \MpchInv. The magnitude of deviations from GR as fostered by these MG models is typically $\mathcal{O}(10\%)$ in these regimes. Therefore reaching 5\% PS modeling is enough for forecasting constraints on modern-era cosmological observables.

\end{abstract}

\maketitle


\section{\label{sec:intro}Introduction}

The standard model of cosmology, the Lambda-Cold-Dark-Matter (\lcdm{}), has been remarkably well-tested observationally in the last two decades. Presently, it is our best approximation of the real Universe \citep{bao_sdss_2005, Planck2018,bbn_review,sdss4_highz,sdss2_snls_de}. The precise observations of the Cosmic Microwave Background (CMB) radiation \cite{Planck2018, WMAP9}, large-scale galaxy clustering \cite{sdss4_highz,bao_2df_2005,wiggleZ_DES,vipers_2017}, and the abundance of massive galaxy clusters \cite{planck13_sz} among others, form a long list where the standard cosmological model predictions are successful. 

So far, the bulk of these \lcdm{} observational  tests concerns the linear regime, the large scales, and/or early times. But it is the mildly non-linear and fully non-linear density fluctuation regimes where the vast majority of the modifications to \lcdm{} are expected to deviate significantly from the standard model predictions \cite{bean_constraints_growth, MG_SIGNATURES_HIERARCHICAL_CLUSTERING, MGCOLA, ECOSMOG_2, speeding_vainshtein, nonlinear_pk_koyama}. It is also in this regime, stretching usually from hundreds down to a few Megaparsecs, where the present and upcoming cosmological surveys like \textsc{desi} \cite{DESI_2013}, \textsc{lsst} \cite{LSST_2019},  and \textsc{euclid} \cite{euclid_2011} aim to measure various statistics concerning the large-scale structure to a percent level accuracy. As a result, with the influx of data from these surveys, the level of the statistical errors can get so small that the measurements start to be more sensitive to systematic effects. If both the new level of accuracy of base-level predictions as well as the control of the known systematics will be successfully implemented, these new large-scale surveys will yield new unprecedentedly accurate estimates and constraints on cosmological parameters: like the DE equation of state, the growth rate of structure, or parameters quantifying possible departures from the standard General Relativity (GR)-based structure formation scenario.

In this context, one of the most useful and widely used theoretical quantities is the power spectrum (PS) of density fluctuations, \pk{}. This statistic generally characterizes the properties of large-scale structures across vast cosmological epochs and scales. Not only it can be used as an end-goal model prediction on its own, but it is also a basic quantity that is used to model and forecast a number of other useful LSS observables, including galaxy clustering measures, cluster abundance, weak-lensing shear and convergence,  the amplitude of the bulk peculiar galaxy motions, and many others \cite{bao_sdss_2005,wl_2013_1,2019_HSC, LSST_2019,sdss4_highz,des_y1_prl,6df_pv}.

Since the PS forms a basis for the predictions of many cosmological LSS observational statistics, the accuracy, and scales to which we know the input PS limits our abilities to forecast the derived observables. Thus, obtaining robust estimates of the PS beyond the linear theory regime (\ie{} scales of $k > 0.1$ \MpchInv) became of paramount importance to modern cosmology, and has been a subject of extensive effort in recent years \cite{Cooray_Sheth_2002,halofit_original,mead_hm_2015,emulator_coyote}. 
A classical approach is to either directly use the results of a number of N-body simulations of LSS evolution to predict PS, or use the simulation results for calibration of more or less approximate models \citep{Cooray_Sheth_2002, halofit_original,halofit_improved,mead_hm_2015,ngenhalofit}. 
Recently, machine-learning-based-emulators are also employed for computing non-linear PS \cite[\eg][]{emulator_coyote,emulator_BACCO}. This approach especially depends on the growing computational power.

In recent years, the progress in modeling the PS has been truly significant. The resulting current state-of-the-art PS models for  \lcdm{} are already, or close to, attaining sub-percent accuracy in the non-linear regime, as required for the success of the cosmological tests offered by the incoming big survey data. However, this amazing progress has been mostly limited to the \lcdm{} alone. When it comes to many interesting extensions and modifications of the standard model, such as the whole family of beyond-GR Modified Gravity (MG) scenarios, the current accuracy, and versatility of PS modeling is still very much lacking. The reasons for this are both higher theoretical complications of such models, and their increased levels of non-linearity \cite{Oyaizu_2008_2, ALAM2020_ELEPHANT, HS_PPF_S007,nonlinear_pk_koyama, react_2019, halomod_fR}. For MG models, N-body simulations play an even more important role in fully assessing the effect of the fifth force, and are crucial for disentangling pure MG effects from the standard GR-based scenarios 
\cite{baldi_degeneracies,mg_code_comparison, MG_GADGET_2013}. This is connected with the richer phenomenology of such models \cite{mg_cosmo,gravity_allscales_baker,cosmological_tests_mg, Nojiri2017_MGreview,cosmological_tests_ferreira,fR_gravity_theories, Tessa_Novel_Probes,beyond_gr_review,de_vs_mg,beyond_lcdm}. Given the fact that the MG simulations are usually many times more expensive than the standard \lcdm{} case \cite{speeding_vainshtein,ECOSMOG_2}, it becomes computationally prohibitive to obtain simulation libraries of the same volume and precision for MG, as is possible for \lcdm{}. However, such libraries are necessary to be applied to the proven state-of-the-art emulating or fitting methods to achieve the same precision, and success in modeling MG effects, as we have for the case of \lcdm{}.

In this work, we attempt to remedy the deficit of accurate MG PS modeling. To circumnavigate the problem of prohibitively expensive MG simulations, we explore a different approach. Instead of trying to model the absolute MG PS predictions, we take \lcdm{} to always be our baseline, and build a semi-analytical model for the relative MG effects on the \lcdm{} PS. We build our model on the basis  of a more general \textsc{Halo Model (HM)} approach (\cite{seljak2000_hm}, reviewed by \cite[\eg][]{Cooray_Sheth_2002,asgari_hm}). Next, we demonstrate how various degrees of modeling freedom can be calibrated and constrained already by a relatively small library of N-body simulations, to achieve an unprecedented level of PS modeling in the MG scenarios studied here.

There are many models that can be considered beyond-GR structure formation scenarios. Most of the viable, and at the same time cosmologically interesting ones usually involve some extra couplings to the metric in the Einstein-Hilbert action that manifests themselves as additional degrees of freedom (\textit{d.o.f.}). The propagation (gradient) of this \textit {d.o.f.} induces an additional gravitational force component, called as the \textit{fifth force}, which acts on top of the Newtonian gravitational force on the cosmological scales. However, propagation of a significant fifth force both on small galaxy scales, and in the strong field regime is tightly constrained observationally \cite{gr_solarsystem,gr_pulsar,gr_em_gw,gr_gw,s2_smbh,gr_deflection_light}. Thus, only MG models that exhibit some kind of a fifth force screening mechanism, which, as the name suggests, would \textit{screen} the fifth force in these observationally tested regimes are viable MG candidates  \cite{screening_review,cosmological_tests_mg, Brax_screening_mg,astrophysical_tests_screening,beyond_gr_review}.

The clockwork of MG models and their involved screening mechanisms can differ in many ways. From our point of view, however, we can significantly simplify the subject by focusing just on phenomenological effective modifications to the density fluctuations PS. As our test-case models, we choose variants of two popular MG set-ups: namely $f(R)$ \cite{fR_gravity_theories} and nDGP gravity \cite{ndgp_2000}, which will serve as a good representative for their whole respective families. Further in the text, we offer a more detailed description and definitions of these models.

Most of the works that have considered computing the non-linear PS in MG models either rely on simulations \citep{Oyaizu_2008_2, ALAM2020_ELEPHANT}, post-Friedmann (PPF) formalism \cite{HS_PPF_S007}, or perturbation theory focusing on quasi-linear scales \citep{nonlinear_pk_koyama,react_2019}. In Ref. \cite{MG_HMCODE}, the non-linear PS is computed  using the \textsc{hmcode} \cite{mead_hm_2015} for a variety of extensions to the standard cosmological model, including $f(R)$ and nDGP. The level of this prediction is however significantly limited by a number of approximations. For example, a simplified spherical collapse theoretical formalism is used there to estimate DM halo properties. From another perspective, \textsc{mg-halofit} was proposed in \citep{mghalofit_fR} as an extension of standard \halofit{} for $f(R)$ gravity models, but \cite{mghalofit_tessore} showed that the former has limited applicability and accuracy.

The HM formalism has been used to model non-linear MG PS in \citep{fR_nonlinear_structure,hm_chameleon,halomod_fR,hm_galileon,hm_nonlocal_gravity,react_2019,sphericalcollapse_braneworld}, which is mainly based on the theoretical spherical collapse model, and is explicitly solved for each MG variant. In our approach, however, we rely on the calibration of phenomenological components of HM to N-body simulations. An additional strength of our approach is that it is general enough to be quite straightforwardly extended, not only to a wider part of the model parameters space but also, in principle, to other modified structure formation models.

This paper is organized as follows: In Section \ref{sec:mg_mod_sims}, we describe the MG models, numerical data sets, and simulations. In Section \ref{sec:HM_intro}, we elaborate on the HM formalism and describe the empirical halo properties: halo mass function (\ref{subsec:theory_hmf}), halo bias (\ref{subsec:theory_bias}) and halo density profile (\ref{subsec:theory_cM}). In Section \ref{sec:results}, we discuss the results obtained from extending the standard HM predictions to the MG models considered in this work (\ref{subsec:shm_res}), and from our new approach (\ref{subsec:hf_ratio}). In (\ref{subsec:lustre_cola}), we test our approach on another suite of MG simulations, and the final Section \ref{sec:conclusions_discussions} includes our conclusions, discussion, and future work prospects. Details of the Appendices are mentioned in the respective sub-sections.

\section{Modified gravity models, numerical data sets and tools}
\label{sec:mg_mod_sims}

As our main data for calibration of the non-linear PS amplitude, we take the \elephant{} (Extended LEnsing PHysics using ANalytic ray Tracing) suite of N-body simulations \cite{ALAM2020_ELEPHANT}. These simulations provide a good test bed to model the impact of $f(R)$ and nDGP physics on formation of the large-scale structure.

In $f(R)$, the fifth force is manifested as a result of additional degrees of freedom from the interaction between an auxiliary scalar field (or \textit{scalaron}) and matter. This additional force appears as a non-linear function of the Ricci scalar, $R$ in the Einstein-Hilbert action, hence the term $f(R)$. We work with the Hu-Sawicki form of $f(R)$ gravity, where \textit{Chameleon screening} screens the fifth force \cite{HS_fR_2007}. In this screening, the scalaron becomes very massive in the high curvature (and high matter density) regimes, and the fifth force exponentially decays above the length scale determined by the inverse of the mass of the scalaron. This length scale is termed the Compton wavelength. As a result of this decay, the scalar interaction diminishes above the Compton wavelength, and GR is recovered \cite{Khoury2003PRD}. 

In the nDGP  model, gravity, unlike other standard forces, mediates from 4D brane to 5D Minkowski spacetime \cite{ndgp_2000,ndgp_2002}. In this model, the scalar is identified as the brane-bending mode which describes the deformation of the 4D brane in the 5D bulk spacetime. The brane bending mode has a second-order term in the equation of motion. On small scales, this term dominates over the linear term. As a result, the coupling between the scalar field and matter is suppressed, and the solutions for metric perturbations approach GR. This is referred to as the \textit{Vainshtein screening} \cite{Vainshtein_1972}. 

In \elephant{}, along with \lcdm{}, two $f(R)$ variants have been employed, with their free parameter $|f_{R0}|$ (the strength of the scalar field today), taken to be 10$^{-6}$ and 10$^{-5}$ (increasing order of deviation from \lcdm) dubbed as F6 and F5, respectively. For nDGP gravity, we have two variants with the model parameter r$_c$H$_0$ = 5 and 1 (which is the dimensionless crossing-over scale characterizing transition from 4D to 5D gravity), marked consequently as N5 and N1, respectively.

The simulations were run from $z_\text{ini}= 49$ to $z_\text{fin} = 0$ employing the \textsc{ecosmog} code \citep{ECOSMOG_1,ECOSMOG_2,ECOSMOG_V_1}, each using $1024^3$ $N$-body particles in a cubic box of a size 1024 \Mpch. The mass of a single particle is $m_p= 7.798 \times 10^{10}$ \msunh{}, and the comoving force resolution is $\varepsilon=15$ \kpch. Each set of simulations has five independent realizations, evolved from the same set of initial conditions. The cosmological parameters of the fiducial background model are given as $\Omega_{m}$= 0.281 (fractional matter density), $\Omega_{b}$ = 0.046 (fractional baryonic density), $\Omega_{\Lambda}$ = 0.719 (fractional cosmological constant density), $\Omega_{\nu}$ = 0 (relativistic species density), $h$ = 0.697 (dimensionless Hubble constant), $n_{s}$ = 0.971 (primordial spectral index), and $\sigma_{8}$ = 0.842 (power spectrum normalization). These parameters apply to background cosmologies in the simulations of all the gravity models. For further processing, we take simulation snapshots saved at $z = 0, 0.3, 0.5$ and $1$.

As indicated above, the \elephant-suite will be our main calibration data set. To test the accuracy of our PS modeling and the general quality of extrapolation, we also use different N-body data. For these additional tests, we take the MG simulations for F5 and N1, described in \cite{naidoo_cs}. These simulations have background cosmological parameters different from our parent \elephant{} simulations, with $\Omega_m$ = 0.3111, $\Omega_b$ = 0.049, $\Omega_\Lambda$ = 0.6889, $\Omega_{\nu}$ = 0, $h$ = 0.6766, $n_s$ = 0.9665 and $\sigma_8$ = 0.8245. This simulation set is run using \textsc{mg-cola} \cite{MGCOLA} in a 500 \Mpch\ box. For each model, we build an ensemble based on five independent realizations.

Linear matter power spectra, \pklin{}, used in this work were calculated using a modified version of the \textsc{camb} cosmological code \cite{CAMB}, which includes a module implementing both the $f(R)$ and nDGP models. The simulation power spectra, \pksim{}, were computed using \textsc{powmes} \cite{powmes}. In what follows, by $P(k)$ we will be denoting the fully non-linear matter power spectrum, unless indicated otherwise.

\section{Halo model formalism}
\label{sec:HM_intro}

As a baseline prediction and our starting point, we take the \textsc{halo model} (HM) approach. It has been proposed as an attempt to analytically model the variance of density fluctuations into the non-linear regime using the properties and clustering of halos as main input parameters. HM describes the statistics of the density field up to the mildly non-linear regimes (\ie{} $k\simlt 0.5$ \MpchInv). Despite its inferior accuracy compared to heavy N-body simulations, the HM has been successfully used for modeling observables and constraining cosmological parameters \citep{Cooray_Sheth_2002,seljak2000_hm,peacock_hm_2000}. 

In HM, the main presumption is that all contributions to the cosmic density field variance come from the matter collapsed into halos.
This allows for moderately accurate modeling of the non-linear two-point clustering statistics, 
although HM can be used to compute the density field at even higher levels of the $n$-point hierarchy \cite{cooray_hm_higherorder}. 

Following HM, the total matter power spectrum \pkhm{} can be described as a sum of two contributions:
\begin{equation}
\label{eqn:hm}
P(k)_{\text{HM}} = P(k)_{1h} + P(k)_{2h}
\end{equation}
where P$(k)_{1h}$ models the contribution from the matter clustered inside halos (called the \textit{one-halo term}) and P$(k)_{2h}$ is the contribution from clustering of separate halos (the \textit{two-halo term}). In practice, the one-halo term dominates at small scales (\ie{} $ k \simgt 1$\MpchInv) and saturates to a constant value at larger scales, where the two-halo term becomes the dominant component of the power spectrum.

These contributions are further defined as:
\begin{equation}
\label{eqn:pk1h}
P(k)_{1h} = \int_{0}^{\infty} dM |\Tilde{u}(k|M)|^{2} \left(\frac{M}{\bar{\rho}} \right)^{2} n(M) 
\end{equation} 
and
\begin{equation}
\label{eqn:pk2h}
 P(k)_{2h} =   I_{m}^{2}(k) P(k)_{\text{lin}} 
\end{equation}
where,
\begin{equation}
    \label{eqn:integral_p2h}
    I_{m}(k) = \frac{1}{\bar{\rho}}\int_{0}^{\infty} dM |\Tilde{u}(k|M)| M n(M) b(M)
\end{equation}
and $I_{m} \rightarrow 1$ for $k \rightarrow 0$ in order to match the linear theory predictions at large scales. The integrals in \cref{eqn:pk1h} and \cref{eqn:integral_p2h} should in principle cover all possible halo mass ranges, but in practice, some M$_\text{min}$ and M$_\text{max}$ are introduced (these mass limits are discussed in more details in the next sub-sections).

Here, $\bar{\rho}$ corresponds to the mean density of the universe, \pklin{} is the linear theory matter power spectrum, $n(M)$ is the \textit{halo mass function}, and $b(M)$ is the \textit{linear halo bias}. 
The term $|\Tilde{u}(k|M)|$ is the normalized Fourier transform of the internal density profile of a halo of mass $M$, such that $\Tilde{u}(k \rightarrow 0, M) \rightarrow 1$. The above HM building blocks are intrinsically redshift-dependent functions, which, in principle, allows one to obtain HM prediction at any redshift for which the integrands are well-defined.

All the components of the HM can be varied independently from each other, and each specific choice of fitting functions, formulae, or tabulated data creates a unique realization. Thus, HM is a general framework under which one can create many different families of PS models. Motivated by literature and our own studies for each of our cosmological models (\ie{} \lcdm{}, and all MG variants), we find an optimal combination of analytic formulae and fitting functions to describe the input properties of halo mass function, halo bias, and halo concentrations. Below we provide a more detailed description of the particular choices we make. For a quick summary and look-up, we refer the reader to \cref{table:fitting_func} which contains a concise list and references of all the fitting functions for the halo properties used in this work, and for each model.

\input{fitting_func.tex}

\subsection{Halo mass function}
\label{subsec:theory_hmf}

The halo mass function (HMF), $n(M)$, quantifies the number of halos per unit mass per unit comoving volume. The most commonly adopted theoretical formulation of the HMF is via the Extended Press–Schechter (EPS) formalism \citep{PS74_MF, Bond_1991}, in which HMF is given by:
\begin{equation}
    n(M) \equiv \frac{dn}{dM} = \frac{\rho}{M^{2}}F(\sigma) \left| \frac{d \ln \sigma}{d \ln M} \right|
\end{equation}
The \textit{halo multiplicity function}, $\fsig = \nu F(\nu)$ denotes the fraction of matter collapsed into halos, in a logarithmic bin around the peak height, $\nu = \delta_c(z)/\sigma(M, z)$. Here, $\delta_c(z)$ is the spherical collapse density threshold, and $\sigma(M, z)$ is the linear variance in the density fluctuation field smoothed using a top-hat filter. This scaling relation has been modeled extensively in the literature and it has been shown to be approximately universal across redshifts for \lcdm{} \citep{S01_MF, J01_MF, W06_MF, W13_MF, D16_MF}. In our earlier work \citep{hmf_sg}, we have shown that after simple re-scaling, the $\fsig{}$ in both $f(R)$ and nDGP also exhibits a similar degree of universality as in the \lcdm-case.

Following our previous study, we will model MG HMF as a fractional deviation, $\Delta_{\text{MG}}$ from the \lcdm{} fiducial baseline, $n(\sigma_M)_{\Lambda \text{CDM}}$. We have shown that such an approach allows for achieving quite a good accuracy ($5-10\%$), which also holds for different background cosmologies. However, to obtain such precision, a careful choice of the baseline \lcdm{} HMF model is paramount.

Thus, for our baseline \lcdm{}, we tested various HMF models in the literature (e.g. \cite{S01_MF, J01_MF, W06_MF, T08_MF,A12_MF, W13_MF, D16_MF}), as these functions can in principle be extrapolated to desired halo mass ranges. We found that the fitting function proposed in  Watson et al. 2013 \cite[][hereafter W13]{W13_MF} proved to be optimal for HM power-spectrum forecasting. Therefore, we used W13 for our \lcdm{} HMF computations.

For completeness, we now recall the essential steps of Ref. \cite{hmf_sg}. Here the target MG HMF is modeled as:
\begin{equation}
    \label{eqn:delta_hmf}
    n(\sigma_M)_{\text{MG}}=\Delta_{\text{MG}}(\sigma_M)\cdot n(\sigma_M)_{\Lambda \text{CDM}}\,\,,
\end{equation}
where $\sigma_M\equiv\sigma(M)$ is simply the linear mass variance at the Lagrangian top-hat halo mass scale, $M$. 

For $f(R)$ gravity models, the fractional deviation fit is expressed as:
\begin{equation}
 \Delta_{\text{MG}} \equiv \Delta_{f(R)} = 1 + a \exp \left[-\frac{(X-b)^2}{c^2}\right],
\label{eqn:fit_fR_equation}
\end{equation}
$X \equiv \ln(\sigma^{-1})$. Here, $(a, b, c)$ are parameters of the fit that were calibrated using  simulations. They  depend on the variant of $f(R)$ gravity model under consideration. See \cref{table:fitting_func} for the specific values that we use in this work.

For nDGP gravity models:
\begin{equation}
   \Delta_{\text{MG}} \equiv \Delta_{\text{nDGP}} = p + q \arctan{(s \, X+r)}.
\label{eqn:fit_nDGP_equation}
\end{equation}

Here, $X$ is the \textit{re-scaled} mass density variance, $X \equiv \ln(\widetilde{\sigma}^{-1})$, $\widetilde{\sigma} = \sigma/\Xi(z)$. Again, $(p, q, r, s)$ are the parameters of the fit, whose values are determined by the variant of the nDGP gravity model.

The resolution of our simulations allowed us to probe only intermediate- and large-mass halos to compute the HMF. In this mass regime, HMF in MG increases w.r.t.\ \lcdm{}, as small-mass halos accrete matter and merge faster to form larger structures. However, this enhanced structure formation at large halo mass-end is happening at the expense of the abundance of smaller halos used up in this process \citep[see \eg][]{HMF_REBEL,est_fR, Falck2015,hmf_sg}. Thus, we can expect that there should be a simultaneous decrease in the number of small-mass halos in the MG models when compared to \lcdm{}.

Equation (\ref{eqn:fit_nDGP_equation}) for nDGP allows the possibility of $\Delta_{\text{nDGP}} < 1$ for small mass halos. However, our fit for $\Delta_{f(R)}$ is never below 1. 
To admit for low-mass halo deficiency also in the $f(R)$, we impose an artificial decrease in $f(R)$ HMF for M $< 10^{11}$ \msunh{},  when compared to \lcdm{} results. For low halo masses, we assume that $\Delta_{f(R)}$ is a linear function of $\ln(\sigma^{-1})$, and is given by:
\begin{equation}
\label{eqn:fR_low_mass}
    \Delta_{f(R)} \rightarrow (m \ln(\sigma^{-1})+n) \times \Delta_{f(R)}
\end{equation}
We tested for different combinations of the $(m,n)$ parameters values. The combination $(m, n) = (0.06, 0.99)$ turned-out to be optimal for our both $f(R)$ variants. Thus we use these values in this work. A note of caution is in place here. There is no clear physical justification for our particular choice of both $m$ and $n$, other than that they are providing optimal HM power spectrum predictions. An interested reader can play around and search for a different choice of $(m, n)$. However, the overall impact of the particular  $(m, n)$ choice
on the resulting HM remains small. 

\subsection{Halo bias}
\label{subsec:theory_bias}
The relation between the clustering amplitude of the underlying DM density field and halos is quantified in terms of the linear halo bias relation, $\delta_h(M)=b(M)\delta$. In the context of power spectra, it is convenient to consider the following Fourier-space estimator of the halo bias:
\begin{equation}
\label{eqn:bias_pk}
    \hat{b}(k,M) = \frac{P_{hm}(k, M)}{P(k)} 
\end{equation}

Here, $P_{hm}(k, M)$ is the halo-matter cross power spectrum, and \pk{} is the matter power spectrum. One can find an optimal value of the linear bias by taking a limit, or an average of this estimator at the smallest possible $k$'s. We consider such a power-spectrum-based bias estimator to use results from \elephant{} suite for testing and finding optimal analytic bias formula for HM.

For this purpose, we tested various $b(M)$ fitting functions for \lcdm{} \citep{seljak04_bias,tinker10_bias, Comparat2017_bias_hmf}. Sheth et al. 2001 $b(M)$ \citep[][hereafter S01]{S01_MF} gave the best match to the simulations. Thus, this will be our choice for the $b(M)$ computations in this work. 

In the \cref{app_sec:theo_sim_bias}, we show the performance of S01, both in capturing the ratio of MG $b(M)$ versus \lcdm{} (\cref{fig:bias_compare_mg_lcdm}), and the absolute $b(M)$ relation (\cref{fig:bias_compare_mg}). We find that S01 gives reasonable predictions in both cases. Given that $b(M)$ impacts only the two-halo term, which by construction matches the \pklin{} on large scales, the choice of $b(M)$ does not impact the HM results to a great extent. 

\subsection{Halo density profile: concentration-mass relation}
\label{subsec:theory_cM}

The scale-free nature of structure formation in CDM scenarios results in self-similar density profiles for individual DM halos, which was first pointed out by Navarro, Frenk, and White in \cite[][hereafter NFW]{nfw_1996}. As a result, DM density profiles are re-scaled by a characteristic central density, $\rho_s$, and radial scale, $r_s$, (or mass $M$ and concentration $c(M)$, respectively). The $c(M)$ relation is defined as the ratio of the virial radius, $R_v$ of the halo to $r_s$, and determines the density profile of NFW halos. 

To obtain relatively unbiased and good-quality NFW fits, the simulated halos need to be well-resolved. The convergence of the halo density profile depends on the simulation's force and mass resolution. Thus $c(M)$ can be reliably estimated only for a limited halo mass range, usually for halo with masses corresponding to at least a few $\times$ 10$^3$ particles \citep[see \eg][]{Power2003}. The resolution of the \elephant{} suite allow only for probing the $c(M \geq 10^{13}$\msunh{}). Because of this, we need to resort to the fitting functions for $c(M)$ here.

We use relations proposed in \citep{mitchell_fR} and \citep{mitchell_ndgp} to compute the $c(M)$ relation in $f(R)$ and nDGP gravity models, respectively. In these works, direct NFW fitting was used to compute the halo density profiles, and  functional forms were derived for the ratio $c(M)_{\text{MG}(= f(R), \text{nDGP})}/c(M)_{\Lambda \text{CDM}}$ (refer to \cref{table:fitting_func} for explicit expressions). The MG $c(M)$ can be therefore obtained as a product of this ratio times the concentration-mass relation for \lcdm{}, for which we use the form proposed in \citep{cM_ludlow16}. Considering the ratio instead of absolute MG $c(M)$ would eliminate the leading-order systematic uncertainties coming from the background cosmology.

The authors in \cite{mitchell_fR} proposed functional form for $\log(c(M)_{f(R)}/c(M)_{\Lambda \text{CDM}})$. When expressed as a function of $M_{500}/10^{p_2}$, this ratio is independent of the background scalar field and $z$. The parameter $p_2$ defined in \cite{hmf_fR_cluster}, encapsulates these dependencies, and in turn allows different variants of $f(R)$ gravity model to be studied in a unified way. 

For the case of the nDGP gravity model, in ref. \citep{mitchell_ndgp}, the ratio $c(M)_\text{nDGP}/c(M)_{\Lambda \text{CDM}}$ is fitted as a decreasing function of $M_{200}$. This fitting also captures the $z$ dependence, hence making the ratio only dependent on the nDGP parameter, r$_c$H$_0$. 

The halo mass range probed in both \citet{mitchell_fR,mitchell_ndgp} is confined to $\geq$ 10$^{12}$ \msunh{}. Therefore, we restrict the use of their  fitting functions to the calibrated mass range, and artificially impose $c(M)_\text{$f(R)$, nDGP} = c(M)_{\Lambda \text{CDM}}$ for $M < 10^{12}$ \msunh{}. 

\section{Results}
\label{sec:results}

In this section, we combine all the HM components to give an analytical prediction for matter overdensity PS. As a reference case to gauge our results against, we always take the PS from \elephant{} simulations. At large scales, the linear perturbation theory gives accurate and reliable predictions both for \lcdm{} and MG PS. Hence, we focus here only on the scales corresponding to mildly and fully non-linear regimes. In practice, we will be interested in the performance of our models for $k \geq 0.1$ \MpchInv. 

\subsection{Halo model predictions for modified gravity}
\label{subsec:shm_res}

We start by testing the standard set-up for HM, which aims to yield a theoretical prediction for the PS amplitude in a given cosmology. For \lcdm{} alone, this approach has at best limited accuracy, since the classical HM fails to accurately capture PS already in the mildly non-linear regime, i.e. $k\simgt 0.2-0.3$ \MpchInv{} \cite{halofit_original, Tinker2005_bias, mead_hm_2015, mead_nl_bias}. Thus, we do not expect that it will perform better in MG cosmologies, which have even richer phenomenology. However, it is still an illustrative exercise, since we will be using this basic HM set-up to obtain much more accurate PS predictions for MG.

Using the inputs of the HMF, $b(M)$, and $c(M)$ relation in their MG versions discussed in the previous section, we compute the resultant power spectra for a number of redshifts. For this, we employ Eqs.\ \eqref{eqn:pk1h}-\eqref{eqn:integral_p2h}, integrating from M$_{\text{min}}$ = 1 \msunh{} to M$_{\text{max}}$ = 10$^{16}$ \msunh{}. We choose a sufficiently broad halo mass range so as to account for the maximum possible halo masses that still have an impact on the resulting PS.

For the integral in Eq.\ \eqref{eqn:integral_p2h} to approach unity at large scales, the bias needs to attain unity when integrated over all the halo masses, i.e.
\begin{equation}
    \label{eqn:bias_unity}
    \frac{1}{\bar{\rho}}\int_{0}^{\infty} b(M) n(M) dM = 1 
\end{equation}
In practice, this integral yields a value below unity, even when the integration is taken over the maximum possible halo mass range. Changing the high mass limit for the integration does not impact the results to a great extent, because on these scales, halos become exponentially rare which makes their contribution to the total power negligible. On the other hand, we expect a significant contribution from the low-mass regime. However, owing to resolution limits, the properties of low mass halos cannot be properly calibrated using simulations. 

Therefore, to add the contribution of the low-mass halos to HM computations, we use the correction proposed in \cite{Schmidt2016_2h_correction,mead_hhm_wl}. This correction adds the contribution of the missing halos to the two-halo term, 
in order to recover \pklin{} at large scales. The correction term is simply yielded by: 
\begin{equation}
    \label{additive_1}
    A = 1- \frac{1}{\bar{\rho}}\int_{M_\text{min}}^{M_\text{max}} b(M) n(M) dM ,
\end{equation}
and it is used as an additive component in the two-halo term:
\begin{equation}
    \label{additive_2}
    C = \frac{A \Tilde{u}(k|\text{M}_\text{min})}{\text{M}_\text{min}}
\end{equation}
Here, $\Tilde{u}(k|\text{M}_\text{min})$ is the normalized Fourier transform of the density profile for the lowest resolved mass M$_\text{min}$. 
Equation (\ref{eqn:pk2h}) is then modified and the resultant two-halo term is given by:
\begin{equation}
    \label{eqn:new_pk2h}
    P(k)_{2h} = P(k)_{\text{lin}}(I_{m}+C)^{2}
\end{equation}

One could instead replace the $P(k)_{2h}$ term with \pklin{}, as the former differs from the latter only for $k \geq 1$ \MpchInv, where already $P(k)_{1h}$ takes over as the dominant contributor. However, for completeness, we use the full above expression for the two-halo term.

\begin{figure}
    \centering
    \includegraphics[width=\columnwidth,keepaspectratio]{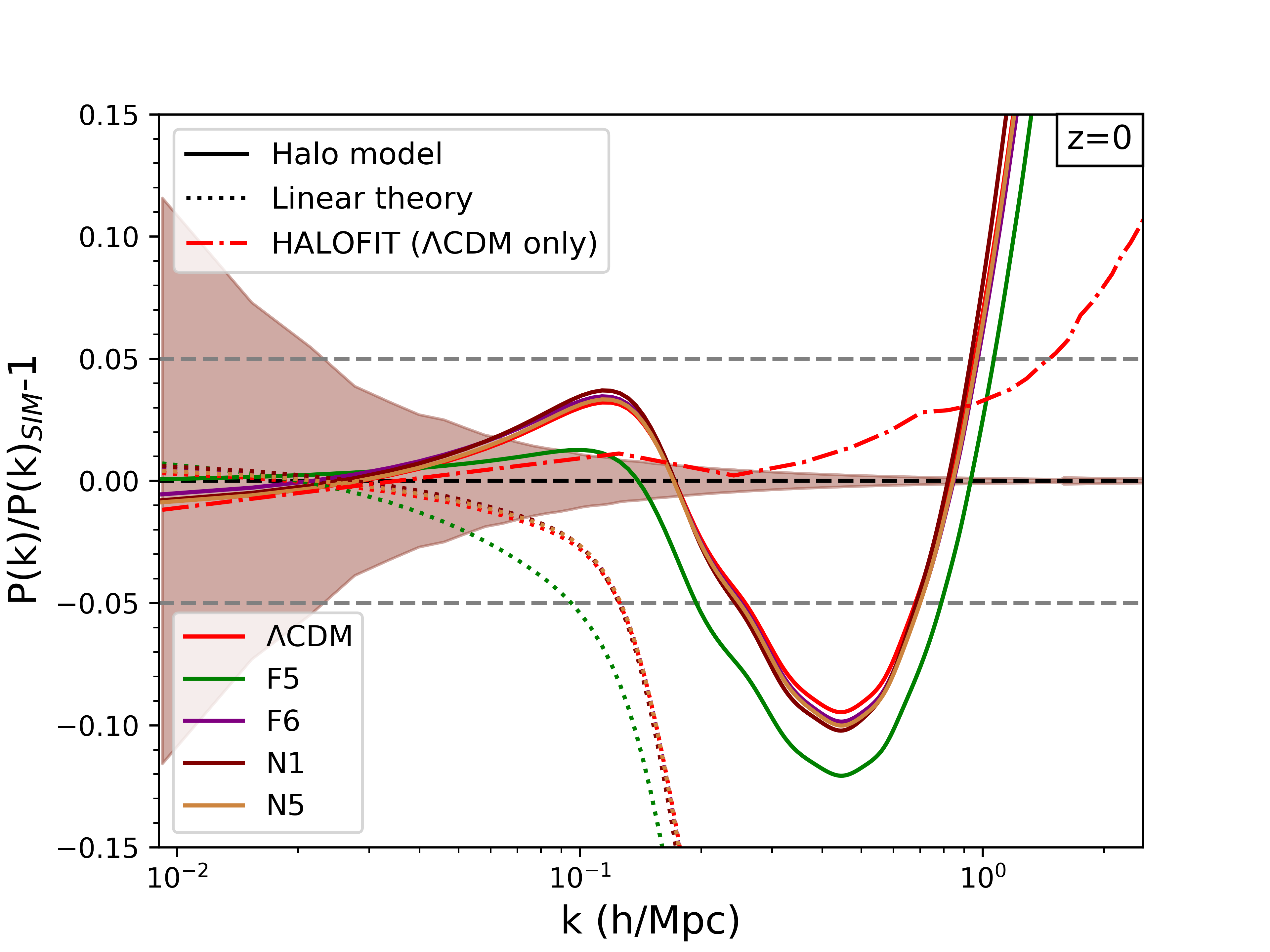}
    \caption{Comparison of the power spectrum from halo-model, \pkhm{} (solid lines), linear theory, \pklin{} (dotted lines) for \lcdm{} and the variants of MG models, and \lcdm{} \halofit{} (red dot-dashed line) with \elephant{} simulations \pksim{}, for $z=0$. The horizontal gray dashed lines correspond to 5\% accuracy regime.}
    \label{fig:pk_compare_shm_sim}
\end{figure}

The results of such direct HM computations for our MG models are illustrated in  \cref{fig:pk_compare_shm_sim}, where we compare \pkhm{} (solid lines), as well as linear theory \pklin{} (dotted lines), with the \elephant{} simulations for all our models at $z=0$. The shaded region corresponds to the uncertainty in the \elephant{} results, which is the inverse of the square root of the number of statistically independent modes contributing to each $k$-bin, and the horizontal dashed lines correspond to the 5\% accuracy regime. The performance of HM in these MG models is similar to the \lcdm{} results and is not much better than the actual linear theory. With respect to the simulation prediction, \pkhm{} gives better than 5\% accuracy for $k \leq 0.2-0.3 $ \MpchInv, and stays within 10\% for $k \leq 0.4-0.5$ \MpchInv. An interesting exception is the F5 $f(R)$ variant, where better than 2\% accuracy is kept all the way to $k\sim0.2$ \MpchInv.

In all the models, we also encounter an under-prediction w.r.t. the simulation results for $k \approx 0.5$ \MpchInv. This is a well-known problem of the HM formalism in \lcdm{}  \citep{mead_hm_2015,mead_nl_bias}, and further propagates to the MG scenarios (also seen for Galileon models in \citep{hm_galileon}). Similar behavior is observed also for other redshifts that our simulations probe, but we do not show them here for brevity.

Our results clearly indicate that HM alone cannot provide modeling accuracy that will be sufficient for the per-cent level accuracy of future LSS surveys. On the other hand, \halofit{}, which is widely used for \lcdm{} PS modeling can perform substantially better. We illustrate this in addition to HM performance in the \cref{fig:pk_compare_shm_sim}. Considering the 5\%-level accuracy for the comparison, \lcdm{} HM can only be used up to $k\leq 0.3$ \MpchInv, while with the \halofit{} one can do much better reaching up to 5\% for $k \leq 1.5$ \MpchInv{}. Thus, we can exploit this much better \halofit{} performance, by using it as a fiducial \lcdm{} baseline, and model only MG-induce deviation form the base-line by the means of HM.

However, noticing the above, the positive result here is that HM can be actually employed to yield predictions for MG power spectra with the same-level accuracy as for \lcdm{}. This is a somewhat surprising result because the standard HM does not include any room for extra MG physics (like the fifth-force and screening). Yet it seems that self-consistent modifications of HMF, $b(M)$, and $c(M)$ are enough to obtain the usual \lcdm{} HM-level predictions also for different MG cosmologies. This is very encouraging, and as we show below 
this can be used as a strong advantage to build an even better and more accurate PS model for MG.

\subsection{An improved model for MG power spectrum} 
\label{subsec:hf_ratio}

\begin{figure*}
\centering
    \includegraphics[width=\linewidth]{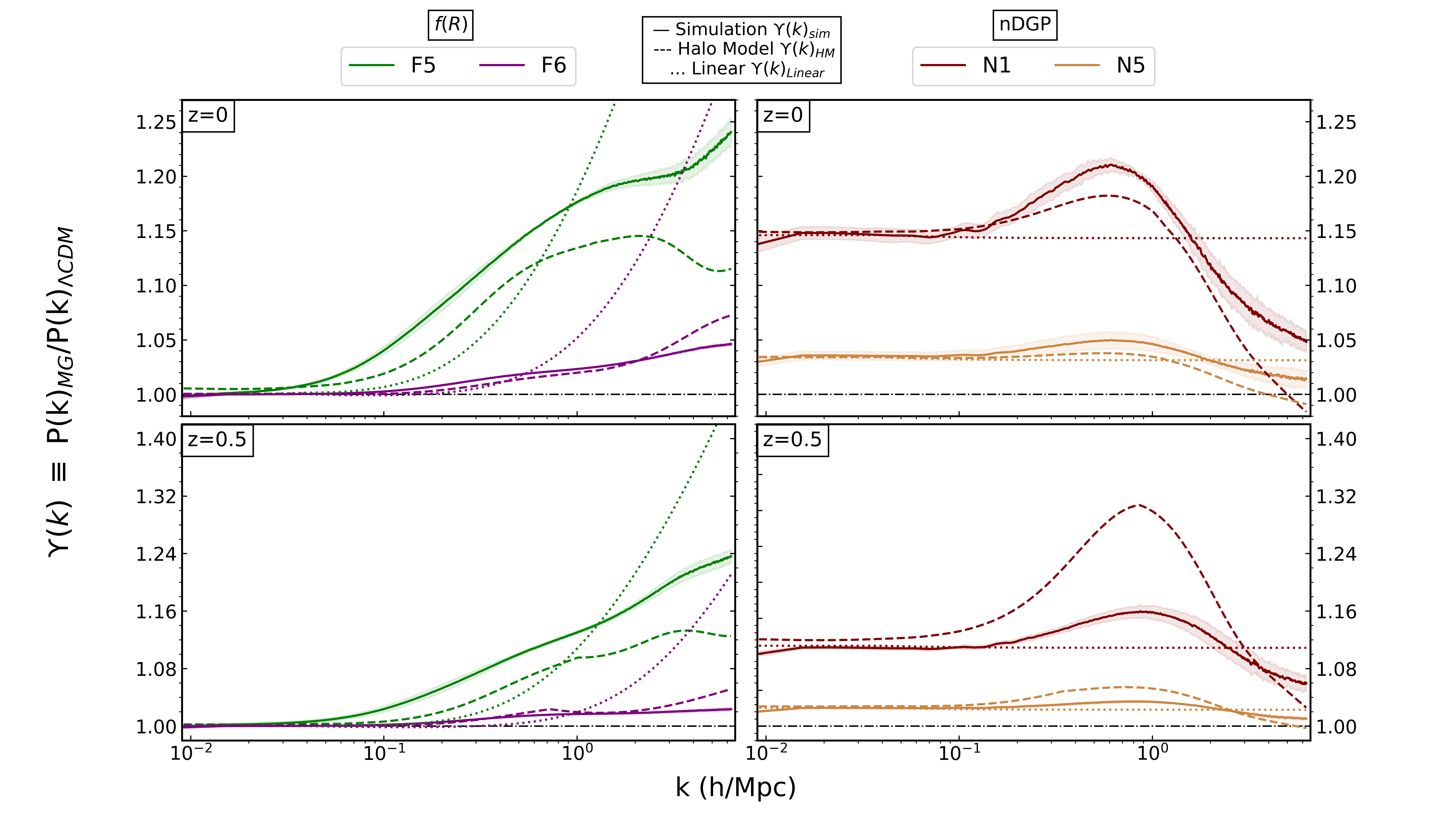}
    \caption{The ratio $\Upsilon(k)$ $\equiv$  $P(k)_{\text{MG}}/P(k)_{\Lambda \text{CDM}}$ obtained from linear theory (dotted lines), \elephant{} simulation (solid lines) and the halo model (dashed lines), at $z=0$ (top panels) and $z = 0.5$ (bottom panels). The left panels correspond to $f(R)$ gravity variants: F5 and F6, and the right panels correspond to nDGP gravity variants: N1 and N5. Shaded regions are the standard deviations obtained for this ratio across five realizations of the simulation box. }
    \label{fig:rato_pk_sim_hm}
\end{figure*}

\begin{figure*}
    \centering 
    \includegraphics[width=0.47\textwidth]{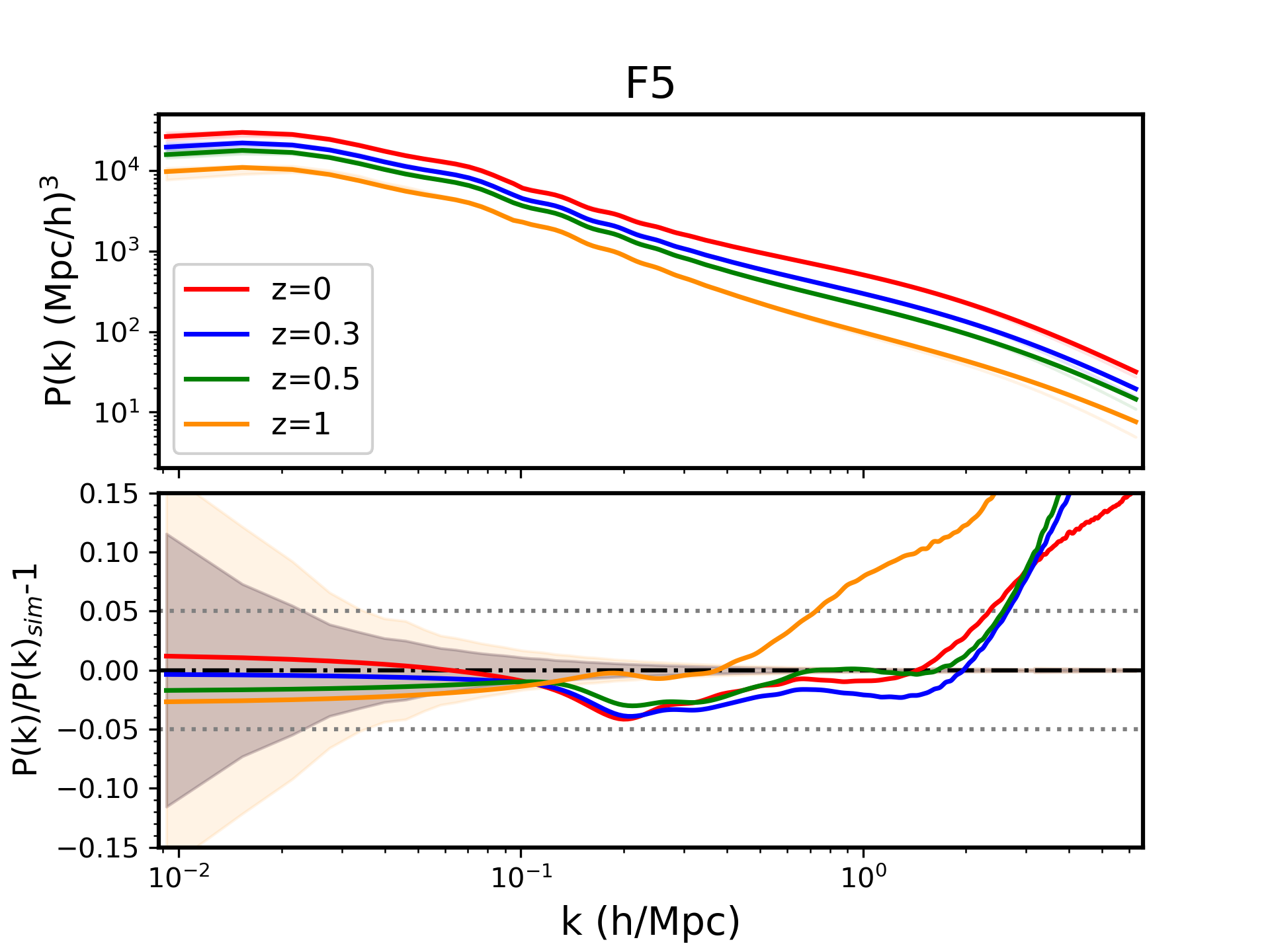}
    \includegraphics[width=0.47\textwidth]{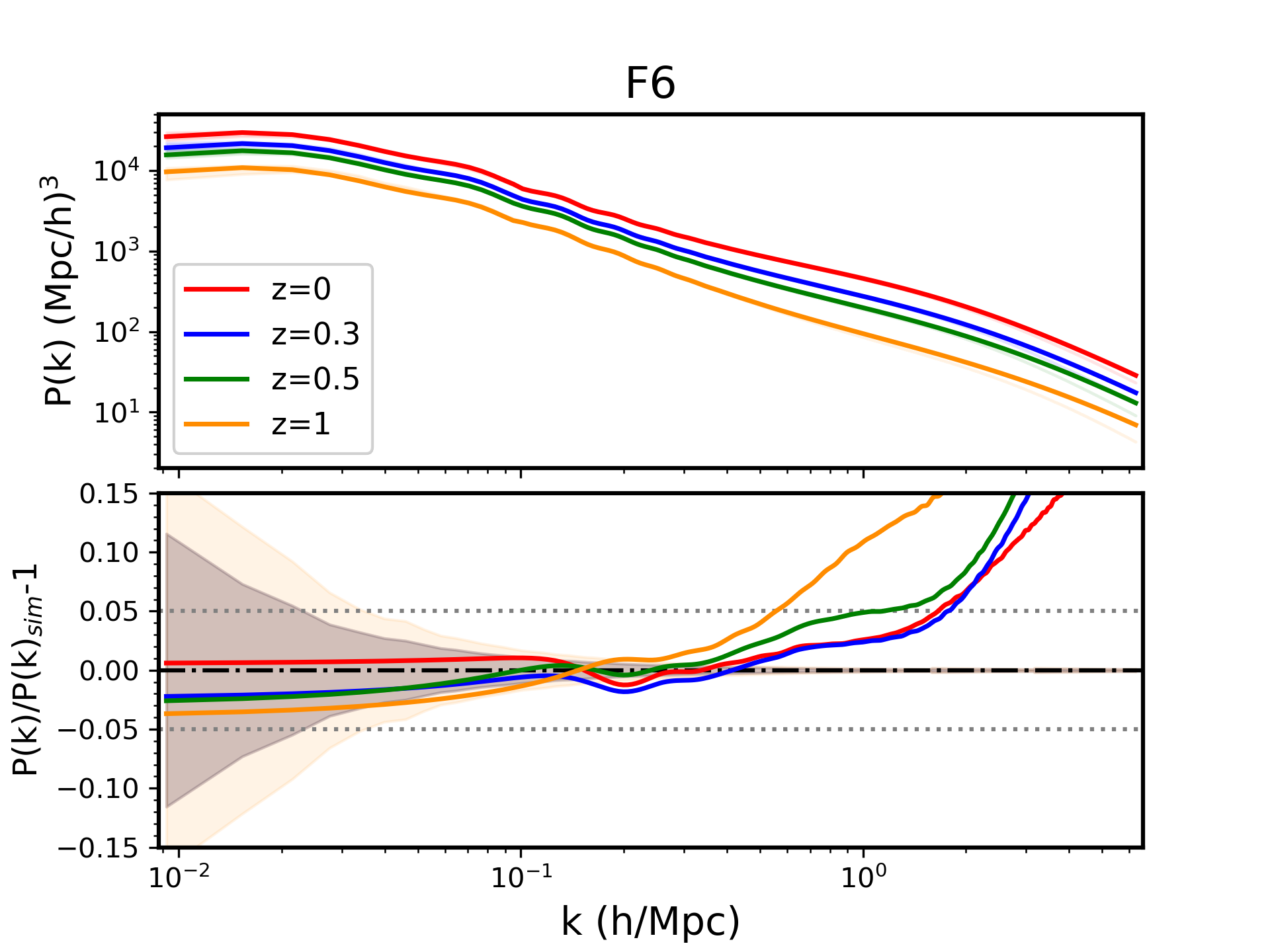}
    \includegraphics[width=0.47\textwidth]{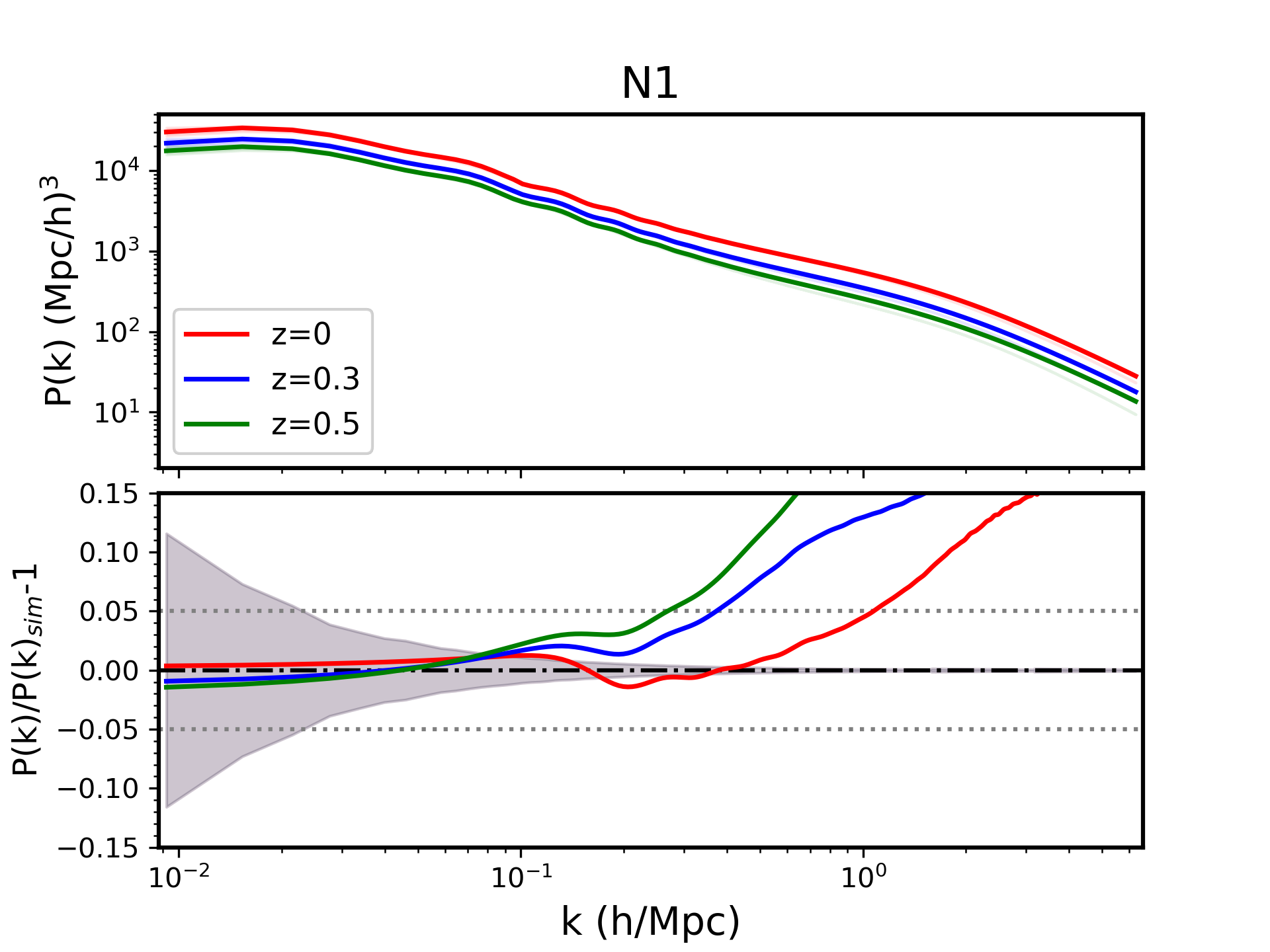}
    \includegraphics[width=0.47\textwidth]{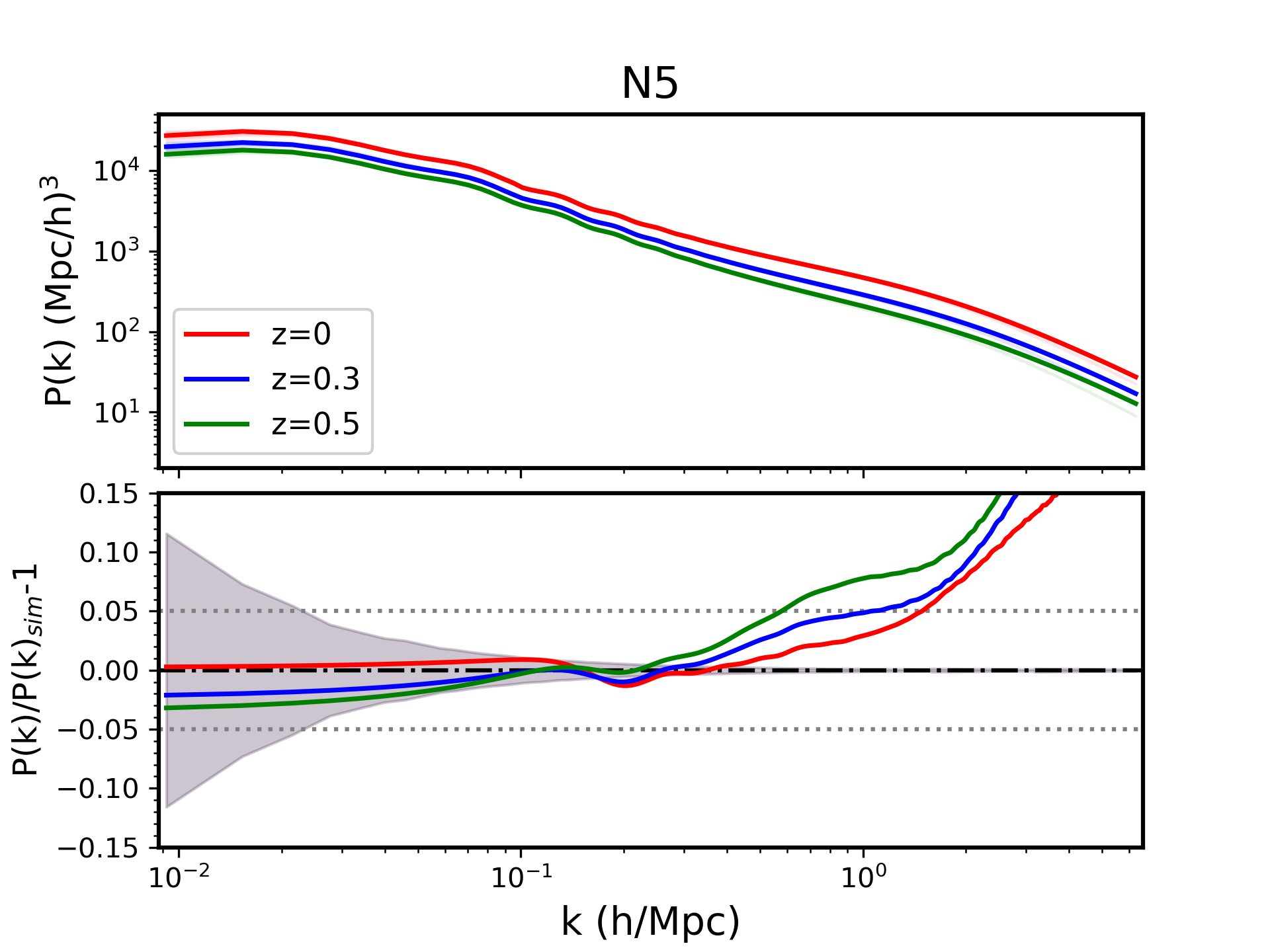}
    \caption{\textit{Top panel:} Matter power spectra obtained from our new approach (\pkmg{} from Eq.~\eqref{eqn:new_pk}) for all the MG variants considered in this work, at redshifts as indicated in the legends. \textit{Bottom panel:} Comparison of \pkmg{}, derived with our new method, with \pksim{}. The shaded region in all the plots corresponds to the uncertainty in the \pksim{}, and the horizontal dotted lines shows 5\% accuracy regime. 
    }
    \label{fig:RATIO_HALOMODEL}
\end{figure*}

In the previous section, we have shown that when HM is applied to model the PS amplitude, it offers  limited accuracy, and is comparable to what can be achieved for the standard \lcdm{}. In this section, we will demonstrate that we can build a much more accurate PS model for MG. This can be realized when we apply HM to estimate the fractional departure from the \lcdm{} baseline, rather than trying to predict the absolute amplitude of PS alone.

Our starting point will be the generic ratio of the MG to \lcdm{} power spectra:
\begin{equation}
    \label{eqn:spectra_reatio_upsylon}
    \Upsilon(k) \equiv P(k)_{\text{MG}}/P(k)_{\Lambda \text{CDM}}\,.
\end{equation}
Here, both numerator and denominator are general terms for MG and \lcdm{} PS respectively. By modeling this ratio, rather than the MG PS itself, we can benefit from a number of properties, namely: (i) the dependence on the background cosmological parameters (such as $\Omega_m$, $H_0$, or $\sigma_8$) should cancel out from the ratio to the leading order; and (ii) the scale of significant departure from \lcdm{} (\ie{} from $\Upsilon=1$) is naturally determined in terms of the \lcdm{} baseline, rather than some arbitrary non-linear amplitude or scale.

In \cref{fig:rato_pk_sim_hm}, we compare the ratios $\Upsilon(k)$ estimated from \elephant{} simulations (solid line), linear theory (dotted line), and HM (dashed line), for both $f(R)$ (left panels), and nDGP (right panels), at $z=0$ (top panel) and $z=0.5$ (bottom panel). Naturally, both the simulation and HM results for $\Upsilon(k)$ are expected to match the linear theory prediction at large scales for both models. As we approach smaller scales, the departure from linear predictions increases (namely, linear theory runs away for $f(R)$ and stays constant for nDGP), and trends peculiar to each model emerge. This is a well-known result, which highlights the fact that these family of MG models usually exhibits an increased degree of non-linearity of the density field, owing to both the fifth force and their respective screening mechanisms \citep{nonlinear_pk_koyama,fR_gravity_theories,screening_review, MG_GADGET_2013,Oyaizu_2008_2, Schmidt_2009_ndgp,vel_mat_ps_fR}. For $f(R)$ gravity models, PS approaches \lcdm{} on the large-scales, and we see a monotonic increase in the ratio with $k$ (although slower than what the linear theory would predict). Whereas, for nDGP, $\Upsilon(k)$ enhancement is maximum at the intermediate scales, and this enhancement decreases for large $k$.

A crucial observation from our study is that HM prediction agrees qualitatively with the simulations. We note that the agreement is far from perfect, especially around the peak-like features, but the HM captures the essential shape and scales of the PS ratios.

As mentioned in the previous section, one perennial problem with the HM has been the \textit{‘transition’} region, where both two- and one-halo terms have a similar magnitude, and both contribute equivalently to the predicted signal. In general, the HM under-predicts the strength of clustering in this region, with the exact amount depending on redshift and cosmology \citep{mead_hm_2015}. We also highlight a similar problem with the HM-based MG predictions in \cref{fig:pk_compare_shm_sim} at $k \approx 0.5$ \MpchInv. These scales are also called the \textit{‘quasi-linear'} regime because the evolution of perturbations at these scales is not exactly governed by linear perturbation theory.

For standard \lcdm{}, the inaccuracies of the HM in this transition regime are addressed by devising empirical fitting functions. One of the earliest, yet successful examples was \halofit{} \cite{halofit_original}, which is motivated by the principles of HM, and calibrated using N-body simulations. It was later improved, in particular by \citep{halofit_improved} who updated its fitting functions from higher resolution simulations and ameliorated the modeling for dark energy cosmologies. Methods and prescriptions to predict the non-linear PS in \lcdm{} are numerous, but in this work we will use \halofit{} as it is sufficiently accurate for our purposes.

Having seen that the ratio $\Upsilon(k)$ between HM-derived PS for MG and \lcdm{}, $\Upsilon(k)_{\text{HM}}$ correctly captures the simulation trends, we propose to use it to obtain the fully non-linear PS in MG. This is done by multiplying $\Upsilon_\mathrm{HM}$ with an accurate model for the \lcdm{} baseline $P(k)$. Therefore, we characterize the beyond-\lcdm{} PS (\pkmg{}) as:

\begin{equation}
    P(k)_\text{MG} = \Upsilon(k)_\text{HM} \times P(k)_{\Lambda \text{CDM} }
    \label{eqn:new_pk}
\end{equation}
Here, $\Upsilon(k)_\text{HM} = P(k)_{\text{MG, HM}}/P(k)_{\Lambda \text{CDM, HM}}$. Both the numerator and the denominator terms are obtained using inputs from \cref{table:fitting_func}. 

In this prescription, \pkmg{} and $P(k)_{\Lambda \text{CDM}}$ are different from \cref{eqn:spectra_reatio_upsylon}. Here, \pkmg{} is the main quantity of focus that we compute in this work, and $P(k)_{\Lambda \text{CDM}}$ is the non-linear \lcdm\ power spectrum, for which we take the \halofit{} predictions using the parameters of a given background cosmology.

The results of applying our proposed methodology are illustrated in \cref{fig:RATIO_HALOMODEL}, where we plot \pkmg{} obtained using $\Upsilon(k)_{\text{HM}}$ multiplied by the \halofit{} \lcdm{}-baseline. The top panels present the power spectra directly: \pksim{} from \elephant{} (dots), and \pkmg{} derived with Eq.~\eqref{eqn:new_pk} (solid lines). In the bottom panels, we show departures of thus-obtained \pkmg{} from \pksim{} treated as reference. These new results, when compared with the standard HM predictions from \cref{fig:pk_compare_shm_sim}, clearly perform much better. The standard HM reaches 5\% accuracy only up to $k \leq 0.2-0.3$ \MpchInv. Now, by using HM only for predicting $\Upsilon(k)_{\text{HM}}$, and combining it with \halofit{} \lcdm{}-baseline, we improve the scale at which modeling is accurate within 5\% by an order of magnitude, reaching up to $k \leq 0.5-2.5 $ \MpchInv{} (depending on the model and redshift). We note that the performance of \pkmg{} generally worsens for higher redshifts, but still remains significantly improved when compared to the standard HM.

\begin{figure*}
    \includegraphics[width=0.45\textwidth]{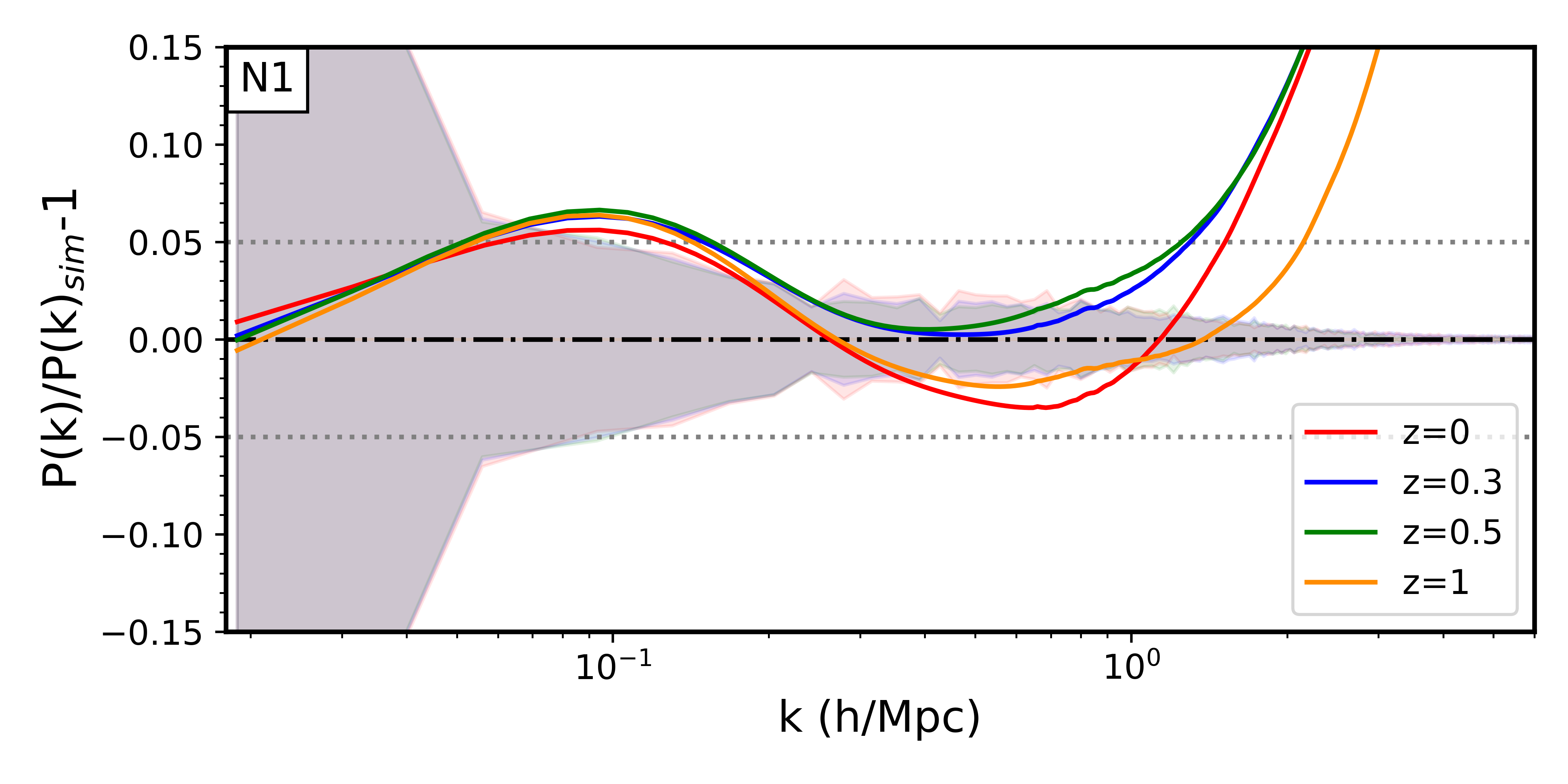}
    \includegraphics[width=0.45\textwidth]{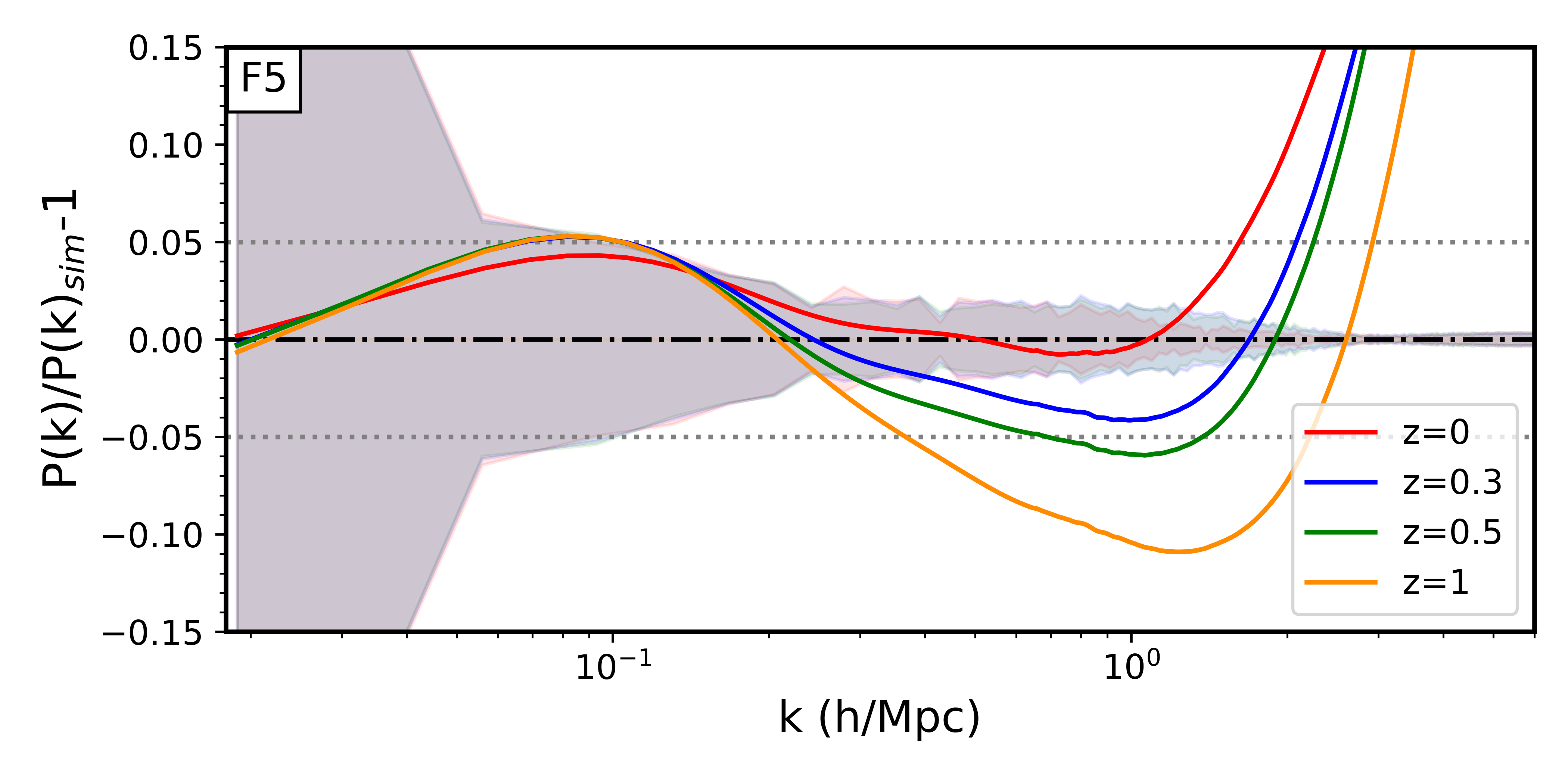}
    \caption{Comparison of our \pkmg{} modeling, with simulation results from \textsc{mg-cola} \cite{MGCOLA, naidoo_cs}, for two MG variants: N1 (left) and F5 (right). The redshifts are as indicated in the legends. Dotted grey lines are the 5\% accuracy regime. The shaded region is the simulation error, which is the standard deviation obtained from five realizations at each redshift.}
    \label{fig:lustre_cola}
\end{figure*}

More generally, the accuracy of \pkmg{} will depend on the user input of baseline $P(k)_{\Lambda \text{CDM}}$. As already mentioned, other approaches are being developed to further improve the limited accuracy of \halofit, especially for models departing from the flat Planck-based \lcdm. We tested one such alternative way of deriving the non-linear \lcdm\ PS, that goes into our \pkmg{} prediction \eqref{eqn:new_pk}: the so-called \hmcode{} \citep{mead_hm_2015,hmcode2020}. The results, detailed in \cref{app_sec:hf_hmcode}, indicate that both \halofit{} and \hmcode{} give similar accuracy, however with different trends at different scales and redshifts.

Given the fact that we have calibrated our MG HM with a limited-resolution \elephant{} simulation suite, it is encouraging that this allowed for already an order-of-magnitude improvement of the scale at which we can obtain accurate PS predictions. Obtaining accurate MG PS into the fully non-linear regime at $k\geq 1$ \MpchInv{} with so straightforward modifications to HM opens up an avenue for even better PS predictions for the MG phenomenology. This could be achieved by incorporating possible improvements to HM that are better informed about the clustering and properties of small halo mass regime in MG.

\subsection{Testing non-linear \pkmg{} beyond ELEPHANT}
\label{subsec:lustre_cola}

In this subsection, we extend our work beyond the \elephant{} simulations to test the performance of our new approach. For this purpose, we consider different N-body simulation runs for the F5 and N1 models, performed using \textsc{mg-cola} \cite{MGCOLA}, and described in \cite{naidoo_cs}. The most important factors for us are that these runs have a different background cosmology than \elephant{} (see \cref{sec:mg_mod_sims}), and were run using different codes. Unlike standard $N$-body approach, these simulations employ the COLA method \cite{COLA}, that can straightforwardly trade accuracy at small-scales in order to gain computational speed without sacrificing accuracy at large scales. On one hand, this approach is much faster than the standard $N$-body, but the price to pay is the approximations made, which do not allow us to use these suite of simulations as the calibration data. Hence, we use these simulations but only as a test-bed.

Here, the HM ingredients were calculated using the same methodology and setup as above, described in \cref{sec:HM_intro} and summarized in \cref{table:fitting_func}. The main difference with respect to Sec.~\ref{subsec:hf_ratio} was that different background cosmological parameters were used in the linear power spectra that go into the particular ingredients of the HM build-up, namely HMF, $b(M)$ and $c(M)$. Everything else, including the halo mass integration ranges for the HM components, were the same as before.

Using the HM outputs and \lcdm{} \halofit{} predictions for the background cosmology of this alternative simulation suite, we computed \pkmg{} (using \cref{eqn:new_pk}). A comparison of our results with the simulation predictions is in \cref{fig:lustre_cola}, for both N1 (left plot) and F5 (right plot). Given the small box size of these simulations ($L$ = 500 \Mpch), we obtain a discrepancy $> 5 \%$ with the simulation predictions on large scales, for $k < 0.1-0.2$ \MpchInv. Now, contrary to the \elephant{} results for N1, our new PS model performs better than before. However, for F5, the performance of our approach decreases with increasing redshift. Overall, we see a similar performance of the new \pkmg{} in both the simulations that we tested, with 5\% accuracy from mildly non-linear to non-linear scales ($k \leq 0.5-2.5$ \MpchInv).

This test with a different simulation and cosmology reassures us that our new approach is a valid technique to compute the non-linear PS in these MG scenarios, and can be successfully extended to simulations and cosmologies beyond our original data that was used for calibration and fitting.

\section{Discussion and conclusion}
\label{sec:conclusions_discussions}

In this work, we combined \textsc{Halo Model} (HM) predictions with an accurate \lcdm{} baseline for building an analytical framework to compute the non-linear power spectrum (PS) in modified gravity (MG) scenarios, where structure formation differs from that in \lcdm{}. For calibration and testing, we used the \elephant{} suite -- a set of $N$-body simulations, which incorporates standard \lcdm{} and two MG models: Hu-Sawicki $f(R)$ and the normal branch of the Dvali-Gabadadze-Porrati braneworld (nDGP). HM has been extensively studied for \lcdm{} \citep{seljak2000_hm,peacock_hm_2000}, and we further extended it to these MG cosmologies. This formalism is advantageous as it is a quick and reliable tool to obtain predictions for statistics of density fields well into the regimes, where linear and perturbation theory fails to reproduce simulation results. 

The HM framework requires the input of three main halo properties: halo mass function (HMF), which quantifies the number density of halos; linear halo bias $b(M)$, describing the relation between halos and the underlying DM density field; and the concentration-mass relation $c(M)$, which describes the internal distribution of mass in halos. For the HM framework, we needed to compute these quantities over large range of halo masses, that go much beyond the range of our simulations. As a result, we relied on fitting functions for the halo properties in these MG scenarios (\cref{table:fitting_func}).

Using these three inputs, we obtained the HM-based predictions, \pkhm{} for these two MG models. We showed that \pkhm{} is within 5-15\% of the simulation results across the $k$-ranges, from $k = 0.01$ to $k = 1$ \MpchInv . However, MG signatures from these models that quantify deviations from GR are typically in itself a factor of a few dozen per cent. Hence, we cannot use HM predictions in its standard form to complement the expected accuracy from future LSS surveys in order to detect these MG signals. Additionally, similar to the case of \lcdm{}, HM also faces the consistent problem of under-prediction of power in the transition regime for both $f(R)$ and nDGP. These scales correspond to $k \approx 0.5$ \MpchInv. 

To get a better PS model, we further investigated, using HM, the relative ratio $\Upsilon(k)_{\text{HM}} = P(k)_{\text{MG, HM}}/P(k)_{\Lambda \text{CDM, HM}}$, instead of employing the absolute PS amplitudes alone. From this, we obtained new analytical PS by taking a product of $\Upsilon(k)_{\text{HM}}$, with the non-linear prediction for \lcdm{}, $P(k)_{\Lambda \text{CDM}}$ (\cref{eqn:new_pk}). For the latter, we used \halofit{} \citep{halofit_original,halofit_improved}, as it has been a successful approach for \lcdm{} to circumvent the HM under-prediction of PS in the intermediate scales, and is widely used to analytically compute non-linear \lcdm{} PS. One could use other approaches for the input non-linear  $P(k)_{\Lambda \text{CDM}}$ to multiply our ratio $\Upsilon_{\text{HM}}(k)$ with, and we tested one of them: \hmcode{} \citep{mead_hm_2015,hmcode2020}. The test with \hmcode{} is shown in \cref{app_sec:hf_hmcode}, and we report that it gives comparable accuracy as \halofit{} for our cosmology.

Using this approach, we significantly improved the accuracy of PS modeling compared to standard HM prediction. For HM, we obtained results within 5\% accuracy with the simulation predictions for $k \leq 0.2-0.3$ \MpchInv. With our new approach, we now matched simulations within this accuracy for $k$ between $0.5-2.5$ \MpchInv, with the performance of the method depending on the MG model and redshift. The $k$-range probed in this work corresponds to the mildly non-linear and the fully non-linear regime: a range of scales crucial to constrain modern era cosmological observables. The sensitivity  of these observables to changes in the matter PS will be very important for making powerful observational cosmological tests of the theory of gravity, or dark energy.

The main advantage of our approach over using simulations is that it is computationally inexpensive. The two main inputs: HM and \halofit{} (or e.g. \hmcode{}) can be flexibly applied to different background cosmologies, whereas, in simulations, we need to perform a new run for each new set of parameters. HM also gives the flexibility of employing different combinations of HMF, $b(M)$ and $c(M)$ that is best suited to probe a particular cosmology, scale, halo mass range, or redshift. 

To test the limits and accuracy of our approach, we applied \cref{eqn:new_pk} to another suite of MG N-body simulations, run with \textsc{mg-cola} \cite{MGCOLA, naidoo_cs}. Using the same fitting functions as described for the \elephant{} simulations, we computed halo properties for the \textsc{mg-cola} cosmology, and in turn \pkhm{}, $\Upsilon(k)_{\text{HM}}$ and \pkmg{}. We compared \pkmg{} with the \pksim{} results, and obtained similar accuracy as with the original data of \elephant{}. For both MG models, \pkmg{} gives consistency with simulation results within  5\% for $k$ between $0.5-2.5$ \MpchInv. However, for the case of F5, overall performance of our approach decreases with $z$.

We need to appreciate that another promising solution for analytical modeling of the MG PS is via the fast and reliable emulation techniques \citep[e.g][]{emulator_main,emulator_main2,emulator_BACCO}. For MG models, emulators have been proposed in e.g. \citep{emulator_pk_fR,emulator_pk_beyond_lcdm,emulator_FORGE,emulator_MGLENS,HM_emulator}. This approach is sophisticated and promising, however, is still in its infancy, and has limitations. For instance, predictions from emulators are confined to the parameter space defined in the starting base grid of the calibrating simulations. In addition, new extensions in emulators (\textit{e.g.} new degrees of freedom, or additional screening mechanisms in MG models) often requires one to substantially adapt the base grid of simulations used to build the emulator, which can in itself be computationally expensive. 

On the other hand, HM potentially provides a simple, physically-motivated semi-analytical picture of the clustering of matter. We showed that HM, in its standard form, can be qualitatively used to predict estimates for MG signatures in cosmological observables which relate to matter perturbations. Furthermore, using HM for modeling the PS ratio $\Upsilon(k)$, and combining it with a high-quality baseline \lcdm{} predictions yields significantly better results. This method is advantageous as contrary to MG scenarios, we have much tighter constraints on \lcdm{} physics, and the field of modeling \lcdm{} PS is much more sophisticated and advanced \citep{mead_hm_2015,halofit_improved,halofit_original,emulator_coyote}. As a result, more precise \lcdm{} results will provide MG PS with similarly improved performance. Here we present our results by incorporating the \halofit{} and \hmcode{} predictions for \lcdm{}. These results in themselves give a percent level of accuracy in both quasi-linear and non-linear regimes.  

In order to further improve HM modeling in the MG variants studied here, we need to probe deeper into the non-linear scales. For this, the behavior of halo density profiles and HMF in both $f(R)$ and nDGP at low halo masses requires deeper investigation, as the full effect of the respective screening mechanisms comes to play in the non-linear regime of gravitational collapse. As mentioned above, the accuracy of $c(M)$ fitting functions for both $f(R)$ and nDGP has not been tested for M$_\text{halo}< 10^{12}$ \msunh{} \citep{mitchell_fR,mitchell_ndgp}. Additionally, we also extrapolated our earlier HMF fits for these MG models \cite{hmf_sg} to small halo mass scales, which are not resolved by our $N$-body simulations (the limit being M$_\text{halo}\lesssim 8 \times 10^{12}$ \msunh{}). Both $c(M)$ and HMF are important ingredients in modeling the one-halo term, which is the dominant non-linear contributor in the HM approach. Such a study will require a completely new set of high-resolution MG N-body simulations, and we plan it as a future project.

We also note that in this work, we focus on modeling only the dark matter PS. At our scales of interest ($k$ between $0.1-2.5$ \MpchInv), PS is not significantly influenced by baryons, as baryonic suppression in PS is of the order of a few percent for $k < 1-5$ \MpchInv{} \cite{alberto_bahamas,mg_lightcone_arnold,chisari_2019}. However, \cite{mead_hhm_wl,hmcode2020,alberto_bahamas} have shown that HM provides the flexibility, which allows it to add additional parameters that can incorporate baryonic effects from hydrodynamical simulation. Accounting for such effects in our MG PS modeling is a significant endeavor, and is well beyond the scope of this work.

The data used here is publicly available on our website\footnote{\url{https://data.cft.edu.pl/UPSILON\_PK/UpsilonPk.tar.gz}}. We provide $\Upsilon(k)_{\text{HM}}$ for a wide range of $z$, from $z=0$ to $z=2$ for each MG model considered in this work. A description of the data set is also enclosed in the directory. Also, the data used to make the figures in this article is available on request to the authors.

\begin{acknowledgments}
We would like to thank Hans A. Winther for kindly providing us with \textsc{mgcamb} version for specific forms of $\mu(a,k)$ and $\gamma(a,k)$ functions implementing our $f(R)$ and nDGP models. The authors also thank Krishna Naidoo for giving his suite of \textsc{mg-cola} simulations that we used as a testbed in our work. We also acknowledge constructive discussions with Marika Asgari, Alberto Acuto and Bartolomeo Fiorini. We also thank the anonymous referee whose comments help improve this manuscript.

Matter power spectrum calculations to compute linear halo bias have been done using \textsc{pylians}\footnote{\url{https://pylians3.readthedocs.io/en/master/}}. This work is supported via the research project ``VErTIGO'' funded 
by the National Science Center, Poland, under agreement no. 2018/30/E/ST9/00698. We also acknowledge the support from the Polish National Science Center within research 
projects no. 2018/31/G/ST9/03388, 2020/39/B/ST9/03494 (WAH \& MB), 2020/38/E/ST9/00395 (MB), and the Polish Ministry of Science and Higher Education (MNiSW) through grant DIR/WK/2018/12. This project also 
benefited from numerical computations performed at the Interdisciplinary Centre for Mathematical and Computational Modeling (ICM), the University of Warsaw under grants no GA67-17 and  GB79-7.

\end{acknowledgments}

\appendix
\section{Comparison of theoretical and simulation halo bias}
\label{app_sec:theo_sim_bias}

\begin{figure*}
    \includegraphics[width=\textwidth]{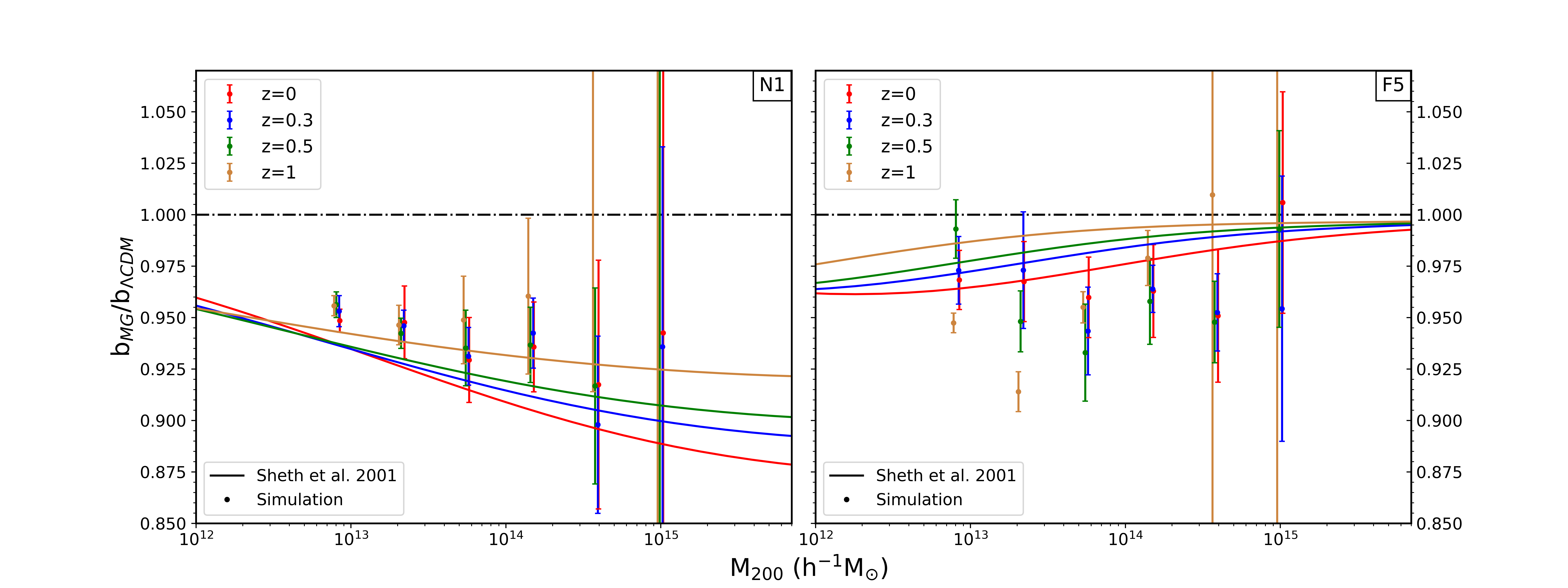}
    \caption{Ratios of halo bias, $b(M)$, between MG and \lcdm{} for N1 (left plot), and F5 (right plot), across range of redshifts as indicated in the legends. Solid lines are the analytical results from Sheth et al. 2001 \cite{S01_MF}, and the respective dots of the same color are from simulations. Error bars illustrate the standard deviation across five realizations of the simulation boxes.}
    \label{fig:bias_compare_mg_lcdm}
\end{figure*}

\begin{figure*}
    \includegraphics[width=0.45\textwidth]{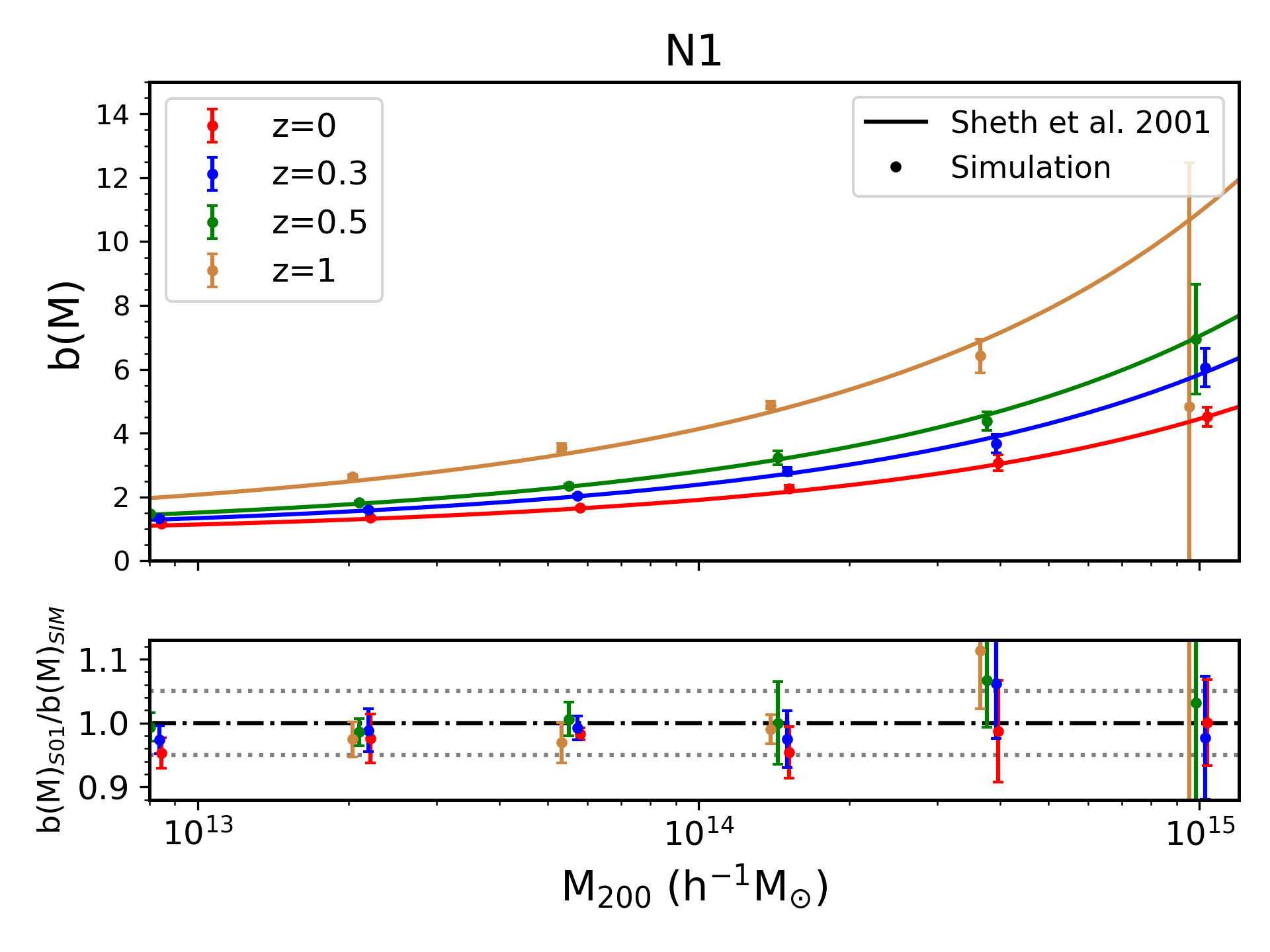}
    \includegraphics[width=0.45\textwidth]{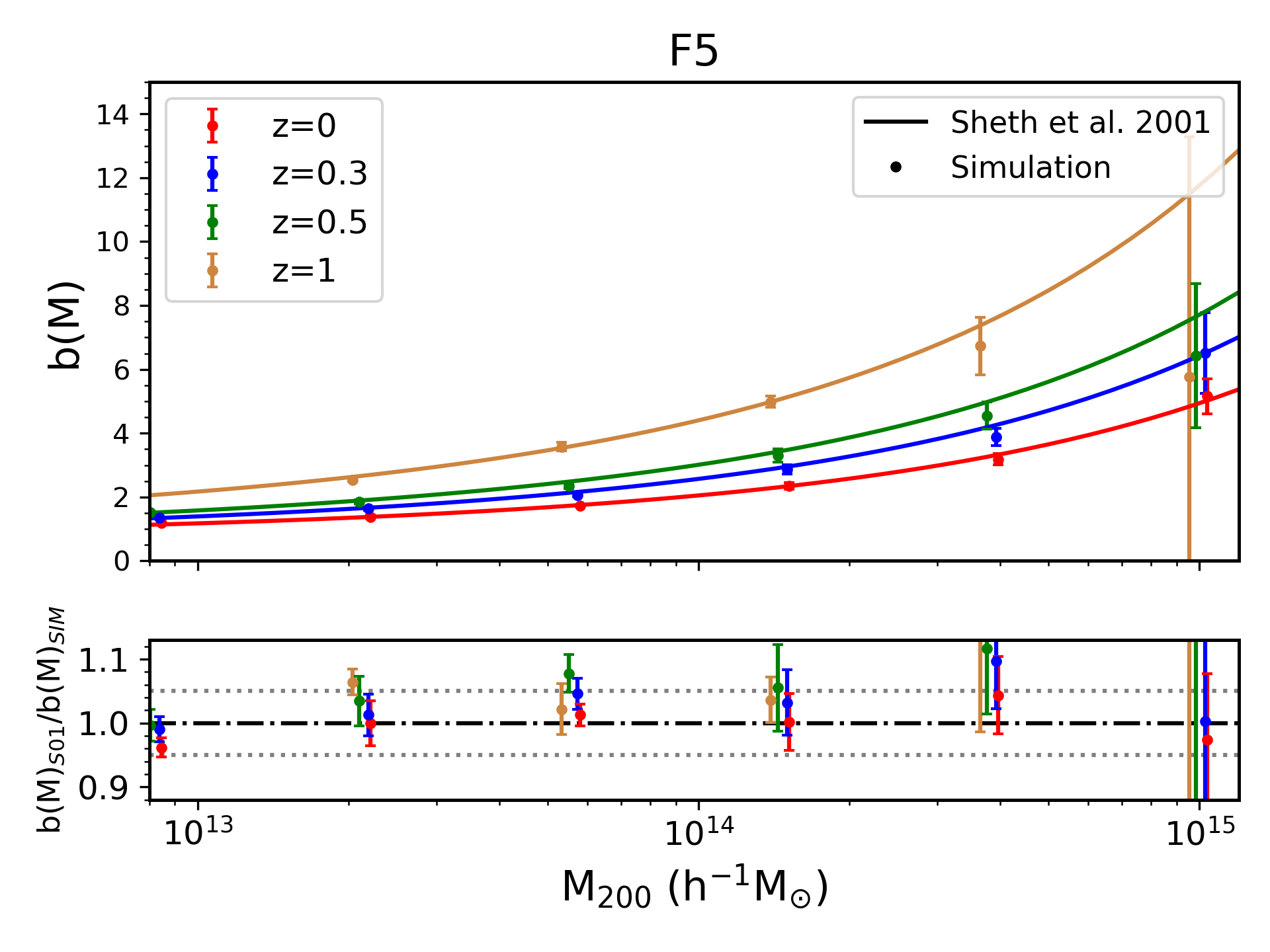}
    \caption{\textit{Top panels:} Linear halo bias, $b(M)$, as a function of halo mass, $M_{200}$ for N1 (left column) and F5 (right column). The solid lines correspond to theoretical Sheth et al. 2001 \citep[S01]{S01_MF} predictions, and the respective dots of the same color are the simulation results obtained using \cref{eqn:bias_pk}. Error bars correspond to the standard deviation across five realizations of the simulation box. \textit{Bottom panels:} Ratio between S01 and simulation linear halo bias predictions. Grey dotted lines are 5\% accuracy regimes.}
    \label{fig:bias_compare_mg}
\end{figure*}

Here we compare the simulation results for the linear halo bias, $b(M)$, with the theoretical predictions from Sheth et al. 2001 \citep[][hereafter S01]{S01_MF}. The formula proposed by S01 is given by:
\begin{equation}
\label{eqn:s01_bias}
\begin{split}
    b(M) = \frac{1}{\sqrt{a}\delta_c(z)}(\sqrt{a}(a\nu^2)+\sqrt{a}b(a\nu^2)^{1-c} \\
    - \frac{(a\nu^2)^c}{(a\nu^2)^c + b(1-c)(1-c/2)} ,
\end{split}
\end{equation}
with the parameters $a = 0.707, b = 0.5$ and $c = 0.6$.

To apply the S01 expression to our MG variants, we used \pklin{} to compute $\sigma(M, z)$, and then $\nu = \delta_c(z)/\sigma(M, z)$, specific to each MG model. For that, we used standard \lcdm{} spherical collapse based $\delta_c$ values.  We stay with the \lcdm{} $\delta_c$ baseline since we have found that using slightly different values suggested for either $f(R)$ \cite{sc_hmf_mg}, or for the nDGP model \cite{sphericalcollapse_braneworld} impacts the final HM results by less than a sub-percent.

The results of our substitution are shown in \cref{fig:bias_compare_mg_lcdm}, where we plot the bias ratios between MG and \lcdm{}, as a function of halo mass, $M_{200}$. Here, we include the two models most departing from \lcdm{}: N1 (left column) and F5 (right column). These variants illustrate the most extreme behavior in $b(M)$ for the two MG models we work with. Points illustrate simulation results, with error bars corresponding to the standard deviation from simulations. For comparison, ratios of S01 predictions for MG and \lcdm{} are also shown, but they are extended outside of the $M_{200}$ ranges probed by our simulations, to show the asymptotic behavior at small and large halo masses. Depending on the redshift and the model, departures in MG $b(M)$ from \lcdm{} can reach up to $\sim10\%$. Contrary to the HMF, MG-induced increase in the strength of gravity lowers the bias, as a result of enhanced DM clustering. Similar trends have also been reported in \citep{Schmidt_Oyaizu_2008_3,sphericalcollapse_braneworld,mg_lightcone_arnold}. The ratios predicted analytically from the S01 framework do not match the simulation amplitudes exactly, but they still qualitatively capture the trends.

Further, in \cref{fig:bias_compare_mg}, we plot the $b(M)$ in these MG models as a function of $M_{200}$. The top panels present absolute $b(M)$ values, while the bottom ones include the ratio between S01 predictions and the simulation-based bias. Here, we can clearly see that the analytical model matches the simulation results within $5-10\%$. This affirms our approach in extending S01 to beyond \lcdm{}, for the $f(R)$ and nDGP models we study. 

As for the HM build-up, we need a bias prescription for a much wider halo mass range than what our simulations cover. This overall consistency between the analytical and simulation results is sufficient for us, and thus we can use the S01 modeling for $b(M)$ also in our MG variants.

\section{Comparison of HALOFIT and HMCODE results}
\label{app_sec:hf_hmcode}

\begin{figure*}
    \includegraphics[width=0.45\textwidth]{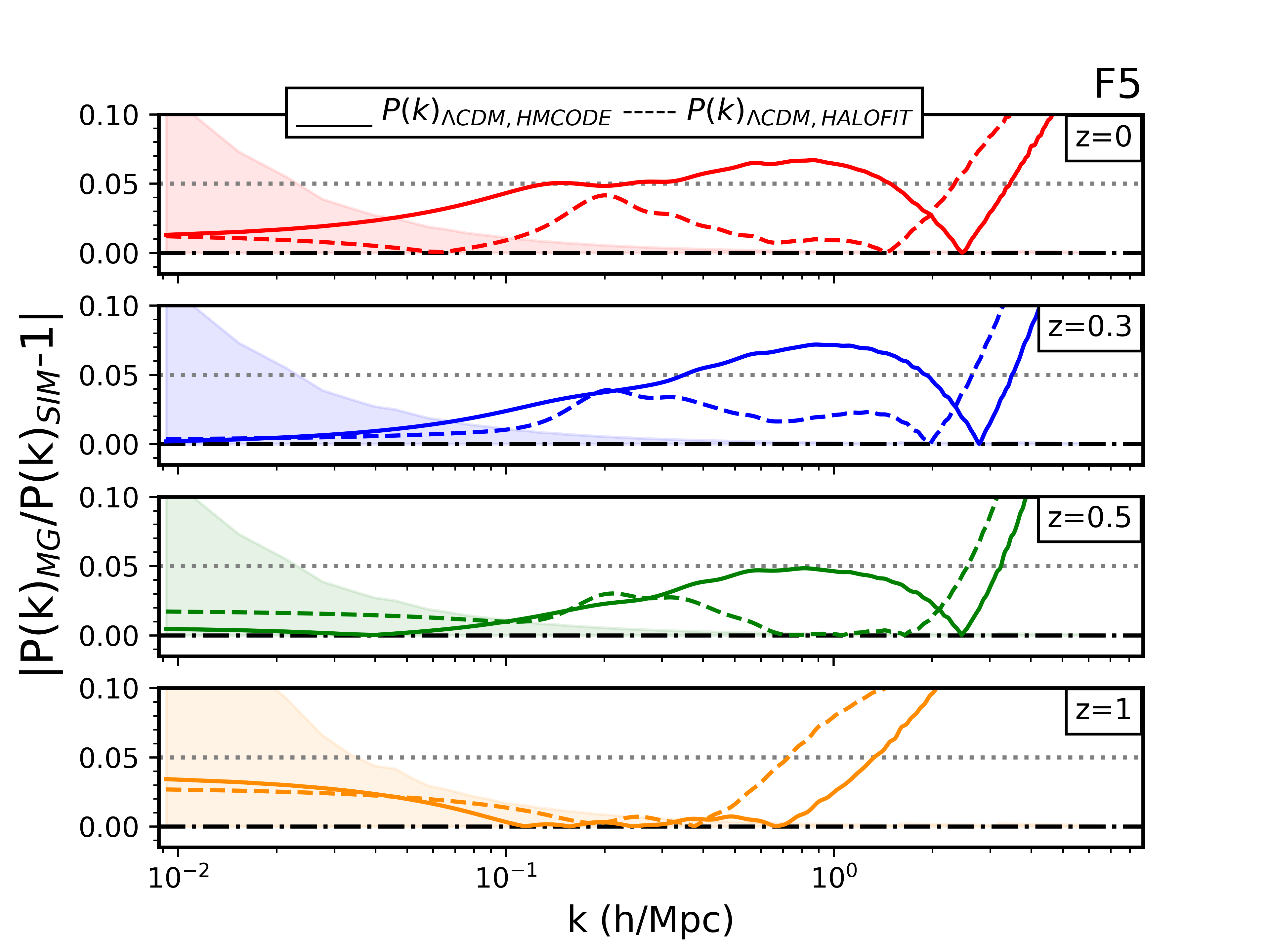}
    \includegraphics[width=0.45\textwidth]{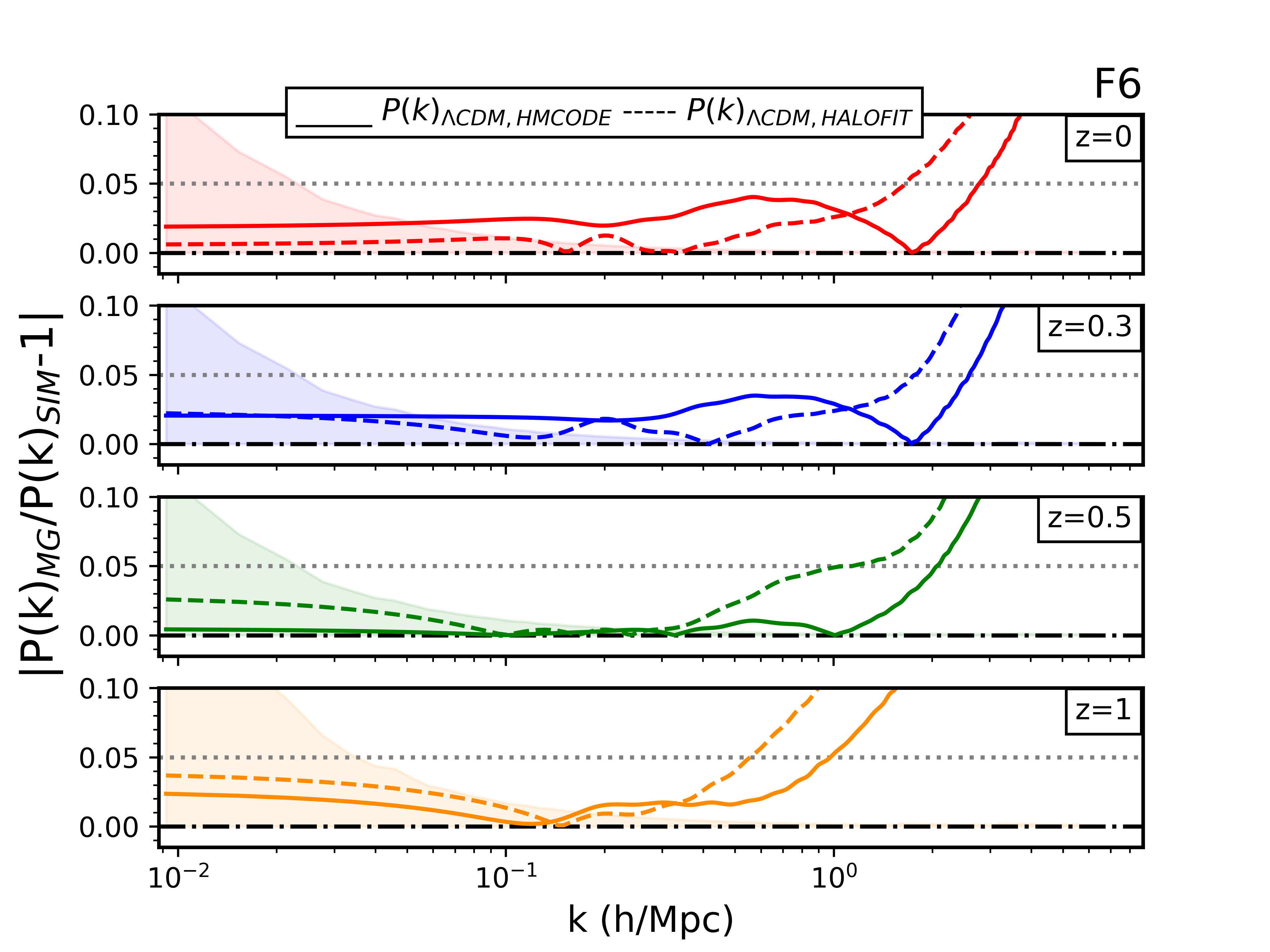}
    \includegraphics[width=0.45\textwidth]{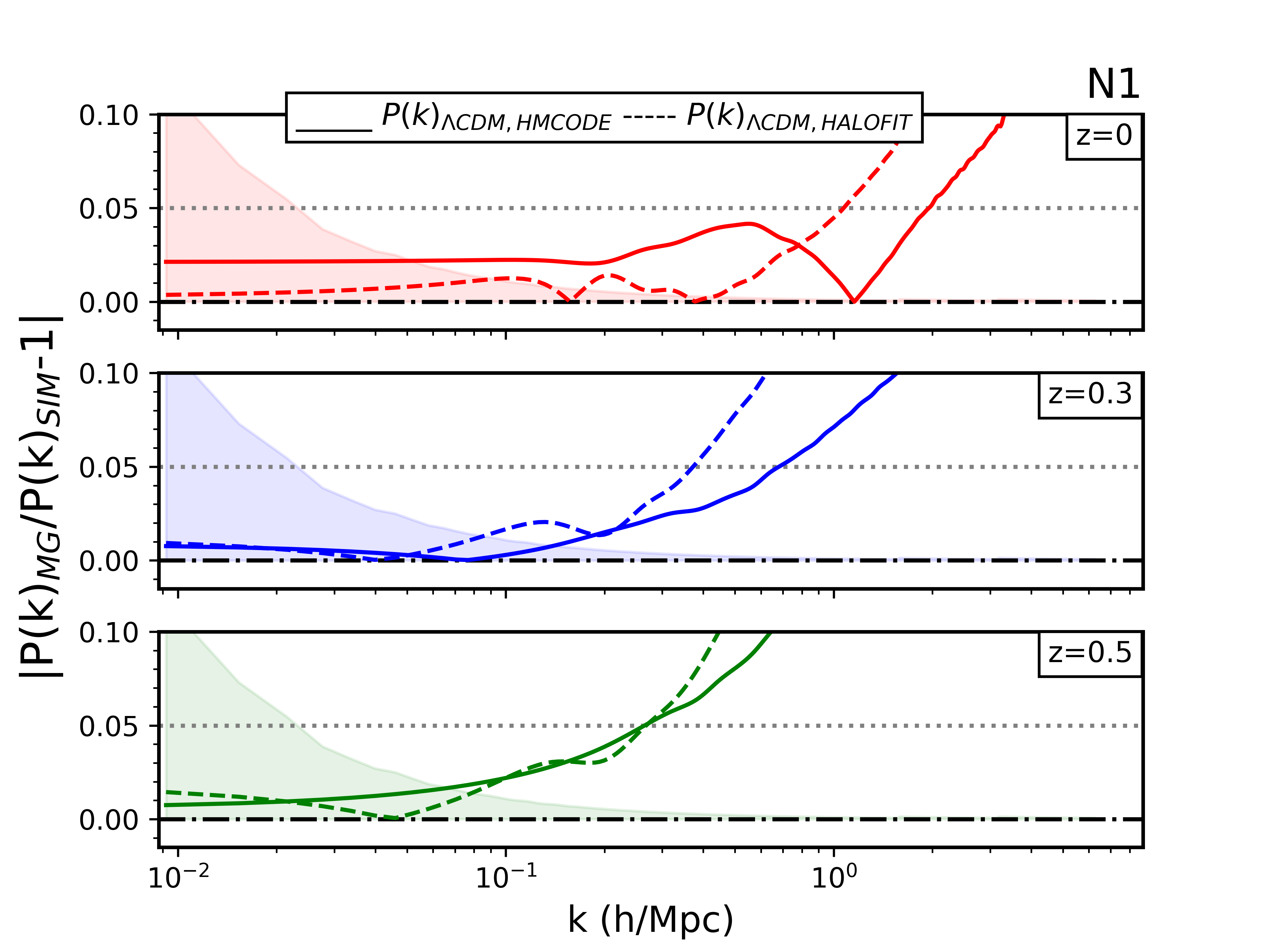}
    \includegraphics[width=0.45\textwidth]{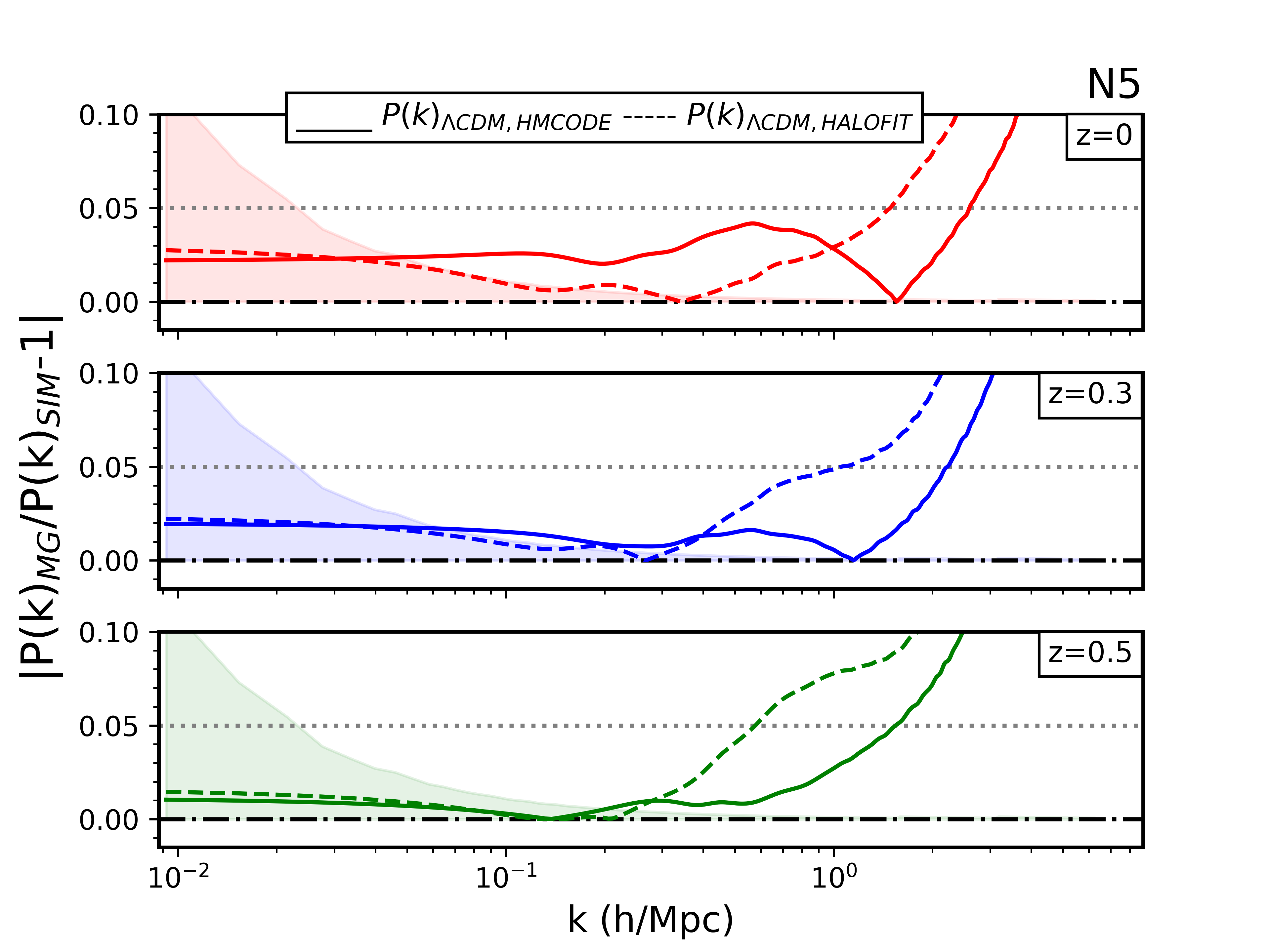}
    \caption{Comparison of \pkmg{} obtained from the input of $P(k)_{\Lambda \text{CDM, HMCODE}}$ (solid lines) and $P(k)_{\Lambda \text{CDM, HALOFIT}}$ (dashed lines), with the \elephant{} simulation results, for a range of redshifts as indicated in the legends. The error contours correspond to the uncertainty in the simulation PS results, and the vertical grey dotted line is the 5\% accuracy regime.}
    \label{fig:hmcode}
\end{figure*}

As discussed in Sec.~\ref{subsec:hf_ratio}, for our baseline PS modeling in MG we multiply the HM-based ratio $\Upsilon(k)$ with \halofit{} PS derived for \lcdm. Here we test our approach for the case where the \lcdm{} PS is obtained from the \hmcode{} \cite{mead_hm_2015} instead.  Similarly to \halofit{}, \hmcode{} is also built on the principles of HM, incorporating however additional corrections in the standard HM build-up, owing to physical constraints. The parameters of the corrections are based on high-resolution simulated \lcdm{} power spectra from the emulator introduced in Ref. \cite{coyote_hmcode}. Here we use the latest `\hmcode-2020' version\footnote{\url{https://github.com/alexander-mead/HMcode}} \cite{hmcode2020}.

We compute \pkmg{} by multiplying $\Upsilon(k)_{\text{HM}}$ with both \halofit{} and \hmcode{} inputs for $P(k)_{\Lambda \text{CDM}}$. Then, in \cref{fig:hmcode}, we compare both predictions with the \elephant{} simulation results. Here we see a similar performance of both the methods, with some exceptions at small scales and high-$z$, where \hmcode{} occasionally performs better. Interestingly, at $z=0$, the \halofit{} framework seems to lead to better results for a range of $k$-scales. We emphasize however that as our simulations were done for one particular set of cosmological parameters, these trends between \halofit{} and \hmcode-based predictions could change for other background cosmologies. In any case, as what we provide is the ratio $\Upsilon(k)_{\text{HM}}$ to be multiplied by the \lcdm{} PS prediction, one can employ any best-fit $P(k)_{\Lambda \text{CDM}}$ for the latter to possibly improve the final accuracy of \pkmg{}.

\bibliography{apssamp}

\end{document}

%% file: fitting_func.tex
\begin{table*}
\centering
\scriptsize

\caption{\label{table:fitting_func} Compilation of the fitting functions used in this work for the halo properties in HM build-up, for both $\Lambda$CDM and MG models.}
\begin{ruledtabular}
\begin{tabular}{c|c|c}
Halo properties &  Fitting functions  & Notes  \\ 
\hline

Halo mass function, HMF &&\\ $\Lambda$CDM : Watson et al. 2013 \cite{W13_MF} & 
$ f(\sigma)_{\Lambda \text{CDM}} = A \left[\left(\frac{\beta}{\sigma}\right)^{\alpha}+1\right]e^{-\gamma/\sigma^{2}} $ &  For MG(=$f(R)$, nDGP),  
\\
&
A = 0.282, $\alpha$ = 2.163, $\beta$ = 1.406 and $\gamma$ = 1.210. 
& $f(\sigma)_{\text{MG}} = \Delta_{\text{MG}} \times f(\sigma)_{\Lambda \text{CDM}}$\\
\hline 
$f(R)$: Gupta et al. 2022 \cite{hmf_sg} & 
 $ \Delta_{f(R)} = 1 + a \exp \left[-\frac{(X-b)^2}{c^2}\right] $ & Additional cut-off expression at \\ & 
For F5: a = 0.230, b = 0.100 and c = 0.360  & low-mass scales for $f(R)$ (\cref{eqn:fR_low_mass})\\&  For F6: a = 0.152, b = -0.583 and c = 0.375  & \\   
& $X \equiv \ln(\sigma^{-1})$ & \\
\hline
nDGP: Gupta et al. 2022 \cite{hmf_sg} & $ \Delta_{\text{nDGP}} = p + q \arctan{(s \, X+r)} $ & \\
 & & $\Xi(z)$: nDGP force enhancement
\\
&
For N1: p = 1.35, q = 0.258, r = 5.12, s = 4.05 & w.r.t. GR \cite{nonlinear_interactions_nDGP}.\\
&
For F6: p = 1.06, q = 0.0470, r = 11.8, s = 4.19 & \\
& $X \equiv \ln(\widetilde{\sigma}^{-1})$, $\widetilde{\sigma} = \sigma/\Xi(z)$ & \\
\hline
\hline

Linear halo bias, $b(M)$ &&\\
All models: Sheth et al. 2001 \cite{S01_MF} & 
$   b(M) = \frac{1}{\sqrt{a}\delta_c(z)}(\sqrt{a}(a\nu^2)+\sqrt{a}b(a\nu^2)^{1-c} $ & This expression has been proposed \\ &
    $- \frac{(a\nu^2)^c}{(a\nu^2)^c + b(1-c)(1-c/2)})$ &  for $\Lambda$CDM. We extrapolated  
\\ & $a = 0.707, b = 0.5$ and $c = 0.6$.
& the relation for MG.\\
\hline
\hline 

Concentration-mass relation, $c(M)$  &&\\
$\Lambda$CDM: Ludlow et al. 2016 \cite{cM_ludlow16} & 

$c(\nu)_{\Lambda \text{CDM}} = c_0 \left(\frac{\nu}{\nu_0} \right)^{-\gamma_1} \left[ 1+ \left(\frac{\nu}{\nu_0} \right)^{1/\beta}\right] ^{-\beta(\gamma_2-\gamma_1)}$
& For MG(=$f(R)$, nDGP),\\ &
$c_0 = 3.395 \times (1+z)^{-0.215}$
& $c(M)_{\text{MG}} = \Delta_{c(M), \text{MG}}\times c(M)_{\Lambda \text{CDM}}$ \\ &
$\beta = 0.307 \times (1+z)^{0.540}$
& \\ &
$\gamma_1 = 0.628 \times (1+z)^{-0.047}$
& \\ &
$\gamma_2 = 0.317 \times (1+z)^{-0.893}$
& \\ &
$\nu_0 = (4.135 - 0.564a^{-1} - 0.210a^{-2}$ 
& \\ &
        $ + 0.0557a^{-3} - 0.00348 a^{-4} ) \times D(z)^{-1}$
& \\
\hline

$f(R)$: Mitchell et al. 2019 \cite{mitchell_fR} &
$ y(x) = \frac{1}{2} \left( \frac{\lambda}{\omega_s}\phi(x')\left[1+erf\left(\frac{\alpha x'}{\sqrt{2}}\right)\right]+\gamma \right)(1-tanh(\omega_t[x+\xi_t]))$ &     \\
&  $ y = log_{10}(\Delta_{c(M), f(R)})$ & \\
&  
$x' = (x-\xi_s)/\omega_s$
&  \\  &
$x = log_{10}(M_{500}/10^{p_{2}})$
&  \\ &
$p_2 = 1.5\log_{10} \left[\frac{\bar{f_R(z)}}{1+z}\right] + 21.64$ \cite{hmf_fR_cluster}
& \\ &
$\lambda = 0.55 \pm 0.18$
&  \\ &
$\xi_s = -0.27 \pm 0.09$
&  For M $\leq 10^{12} M_{\odot}/h$, \\ &
$\omega_s = 1.7 \pm 0.4$
& $c(M)_{f(R), \text{nDGP}} = c(M)_{\Lambda \text{CDM}}$ \\ &
$\alpha = -6.5 \pm 2.4$
&  \\ &
$\gamma = -0.07 \pm 0.04$
&  \\ &
$\omega_t = 1.3 \pm 1.0$
&  \\ &
$\xi_t = 0.1 \pm 0.3 $
&  \\
\cline{1-2}
nDGP: Mitchell et al. 2021 \cite{mitchell_ndgp} & $\Delta_{c(M), \text{nDGP}} = [A - B log_{10}(M_{200}M_{\odot}h^{-1})](H_0r_c)^{-0.71 \pm 0.05} + 1 $ 
&  
\\ & $A = (0.35 \pm 0.01)(H_0r_c)^{-0.71 \pm 0.05}$
&\\ & $B = (0.0302 \pm 0.0008)(H_0 r_c)^{-0.71 \pm 0.05}$

\end{tabular}
\end{ruledtabular}
\end{table*}